\documentclass[11pt]{article}
\usepackage[left=1in,top=1in,right=1in,bottom=1in,head=.1in,nofoot]{geometry}

\usepackage{setspace,url,bm,amsmath} 

\usepackage{titlesec} 
\titlelabel{\thetitle.\quad} 

\usepackage[margin=20pt]{subcaption}

\usepackage{amsmath,amsfonts,amsthm,amssymb}
\usepackage{graphicx}

\usepackage[colorlinks=true,linkcolor=black,citecolor=blue,urlcolor=blue]{hyperref}
\usepackage{threeparttable}

\usepackage{comment}
\theoremstyle{definition}
\newtheorem{assumption}{Assumption}
\newtheorem*{theorem*}{Theorem}
\newtheorem{theorem}{Theorem}
\newtheorem{proposition}{Proposition}
\newtheorem{lemma}{Lemma}
\newtheorem{remark}{Remark}

\newtheorem{corollary}{Corollary}
\newtheorem*{corollary*}{Corollary}

\DeclareFontFamily{U}{mathx}{\hyphenchar\font45}
\DeclareFontShape{U}{mathx}{m}{n}{
      <5> <6> <7> <8> <9> <10>
      <10.95> <12> <14.4> <17.28> <20.74> <24.88>
      mathx10
      }{}
\DeclareSymbolFont{mathx}{U}{mathx}{m}{n}
\DeclareFontSubstitution{U}{mathx}{m}{n}
\DeclareMathAccent{\widecheck}{0}{mathx}{"71}
\DeclareMathAccent{\wideparen}{0}{mathx}{"75}

\usepackage{float}
\usepackage{natbib}
\usepackage{indentfirst}

\def\T{\text{T}}
\def\Var{\text{Var}}

\def\adj{\text{adj}}

\def\ols{\text{ols}}
\def\proj{\text{proj}}

\def\gen{\text{gen}}

\def\taustr{\hat \tau}

\def\taugen{\hat \tau_{\text{gen}} }
\def\tauols{\hat \tau_{\text{ols}} }

\def\tauinter{\tilde \tau_{\text{ols}}}

\def\Lasso{\text{lasso}}
\def\betaLassoT{\hat \beta_{\Lasso}(1)}
\def\betaLassoC{\hat \beta_{\Lasso}(0)}
\def\tauLasso{\hat \tau_{\Lasso}}

\def\betaLassoTs{\hat \beta_{[k] \Lasso}(1)}
\def\betaLassoCs{\hat \beta_{[k] \Lasso}(0)}
\def\tauLassos{\tilde \tau_{\Lasso}}

\def\betaTs{\hat \beta_{[k]}(1)}
\def\betaCs{\hat \beta_{[k]}(0)}

\def\bh{{h}}

\def\r{\varepsilon}

\def\sumi{\sum_{i=1}^{n}}
\def\sumk{\sum_{k=1}^{K} }
\def\sumik{\sum_{i \in [k]}}
\def\nkt{n_{[k]1}}
\def\nkc{n_{[k]0}}
\def\pik{\pi_{[k]}}

\def\pnk{p_{n[k]}}

\def\tauk{ \tau_{[k]}}
\def\taukhat{\hat \tau_{[k]}}
\def\pk{p_{[k]}}

\def\nk{n_{[k]}}

\def\nt{n_1}
\def\nc{n_0}

\def\YkThat{\bar{Y}_{[k]1}}
\def\YkChat{\bar{Y}_{[k]0}}

\def\bx{X}

\def\XkT{\bar{X}_{[k]}}
\def\XkC{\bar{{X}}_{[k]}}
\def\XkThat{\bar{{X}}_{[k]1}}
\def\XkChat{\bar{{X}}_{[k]0}}

\DeclareMathOperator*{\argmin}{arg\,min}

\title{\bf  A general theory of regression adjustment for covariate-adaptive randomization: OLS, Lasso, and beyond}
\date{}

\author{
\small
{
Hanzhong Liu$^{1}$, \ \ Fuyi Tu$^{2}$, \ \ Wei Ma$^{2}$\thanks{\small{Correspondence: \texttt{mawei@ruc.edu.cn}}}
}
\\ \\
{\small $^{1}$ Center for Statistical Science, Department of Industrial Engineering, Tsinghua University, Beijing, China}\\
{\small $^{2}$ Institute of Statistics and Big Data, Renmin University of China, Beijing, China}
}

\begin{document}
\doublespacing

\maketitle


\begin{abstract}

We consider the problem of estimating and inferring treatment effects in randomized experiments. In practice, stratified randomization, or more generally, covariate-adaptive randomization, is routinely used in the design stage to balance the treatment allocations with respect to a few variables that are most relevant to the outcomes. Then,  regression is performed in the analysis stage to adjust  the remaining imbalances to yield more efficient treatment effect estimators. Building upon and unifying the recent results obtained for  ordinary least squares adjusted estimators under covariate-adaptive randomization, this paper presents a general theory of regression adjustment that allows for arbitrary model misspecification and the presence of a large number of baseline covariates. We exemplify the theory on two Lasso-adjusted treatment effect estimators, both of which are optimal in their respective classes. In addition, nonparametric consistent variance estimators are proposed to facilitate valid inferences, which work irrespective of the specific randomization methods used. The robustness and improved efficiency of the proposed estimators are demonstrated through a simulation study and a clinical trial example. This study sheds light on improving treatment effect estimation efficiency by implementing machine learning methods in covariate-adaptive randomized experiments.

\vspace{12pt}
\end{abstract}

\section{Introduction}

Randomization is considered as the gold standard for evaluating treatment effects in interventional studies. The most basic randomization method, simple randomization, allocates the treatment for each experimental unit with a fixed probability. However, under simple randomization, notable imbalances often occur on important baseline covariates. In practice, stratified randomization, or more generally, covariate-adaptive randomization, is often used to balance treatments with respect to a few of the most relevant stratification variables, although other baseline covariates might still be imbalanced  \citep{Rosenberger2008, Liu2020}. An alternative is to use regression to adjust the covariate imbalance in the analysis stage. In this study, we use regression adjustment to robustly and efficiently estimate and infer the treatment effect in covariate-adaptive randomized experiments, allowing the baseline covariates to be high-dimensional.

As in \cite{Ma2020Regression}, covariate-adaptive randomization refers to randomization schemes that tend to balance treatments with respect to discrete baseline covariates, and thus, covers many commonly used randomization methods in all fields of science. In particular, stratified block randomization \citep{Zelen1974} and minimization \citep{Taves1974,Pocock1975} are routinely used in randomized controlled clinical trials, and according to recent surveys,  together accounts for approximately 80\% of the papers published  in leading medical journals  \citep{Lin2015,Ciolino2019}. Please refer to \cite{Hu2012} and \cite{Rosenberger2015} for detailed discussions of covariate-adaptive randomization in clinical trials and \cite{Duflo2007} and \cite{Bruhn2009} for reviews of its use  in development economics.

Recent studies have shown that various ordinary least squares (OLS) regression-adjusted estimators are robust under covariate-adaptive randomization, in that the inferences are valid even if the regression models are arbitrarily misspecified. Thus, these inferences are more flexible than the model-based methods \citep[e.g.,][]{Shao2010,Ma2015} that require a correctly specified data generation process.
Regression adjustment only for stratification variables was thoroughly studied by  \cite{Bugni2018,Bugni2019}.
When considering additional baseline covariates, \cite{Ma2020Regression} and \cite{Ye2020Inference} separately proposed stratum-common and stratum-specific OLS-adjusted treatment effect estimators to further improve efficiency. Both of these estimators are optimal in their own contexts (see Section 4 for more details). These approaches to adjusting covariate imbalances have also been shown to be effective for stratified randomization under a finite-population framework \citep{Liu2019}.

Nevertheless, the common drawback of the above-mentioned methods is that they incorporate only a few baseline covariates for regression adjustment. In the present era of big data, the number of covariates can be very large, even larger than the sample size. For example, in clinical trials, patient history, demographic and disease characteristics, and genetic information are collected at baseline. These high-dimensional covariates, although generally not used in the design stage, may be a valuable source for a more efficient treatment effect estimation. Because the OLS estimators tend to fail in the high-dimensional settings because of  over-fitting, it is conceptually desirable to use penalized regression, such as the Lasso \citep{Tibshirani1996}, to estimate the treatment effects in randomized studies \citep[e.g.,][]{Tsiatis2008, Tian2012}. However, rigorous justification is limited and mainly applicable to simple randomization \citep{Bloniarz2016,Wager2016,Liu2018,Yue2018}. Most relevantly to this paper, \citet{Bloniarz2016} studied a Lasso-adjusted treatment effect estimator under a finite-population framework, which was later extended to other penalized regression-adjusted estimators \citep{Liu2018,Yue2018}.

This paper presents a general theory of regression adjustment for the robust and efficient inference of treatment effects under covariate-adaptive randomization. This study builds upon the recent advances made in OLS-based inference and is adapted to the settings with high-dimensional covariates. We exemplify our theory on two Lasso-adjusted treatment effect estimators. Asymptotic properties are derived under mild conditions, and robust variance estimators are  proposed to facilitate valid inferences. The generality of our theory is at least threefold. First, it applies to both low- and high-dimensional covariates. Second, we allow the linear model to be arbitrarily misspecified. Third, the proposed inference procedures do not depend on specific randomization methods.

\section{Framework and notation}

We follow the framework and notation introduced in \citet{Ma2020Regression}. In a covariate-adaptive randomized experiment with two treatments,  let $A_i$ denote the treatment assignment for unit $i$, $i=1,\dots,n$, which takes the value of one for the treatment and zero for the control. We denote $Y_i(1)$ and $Y_i(0)$ as the potential outcomes  under the treatment and control, respectively. The observed outcome is $Y_i = A_i Y_i(1) + ( 1 - A_i ) Y_i(0)$. The experimental units are stratified into a fixed number of strata based on the baseline variables, such as gender, grade, or location. Let $B_i$ denote the stratum label, which takes values in $\{1,\dots,K\}$, where $K$ is the number of strata. For simplicity, we assume that  the probability of units assigned to each stratum is positive, that is, $\pk = P(B_i = k) > 0$, $i=1,\dots,n$, $k=1,\dots,K$. In additional to the stratification variables, we observe a $p$-dimensional vector of baseline covariates, denoted by $\bx_i = (x_{i1}, \dots, x_{ip})^\T$. We consider a high-dimensional setting in which $p$ tends to infinity as $n$ goes to infinity. We use $[k]$ to index units in stratum $k$ and let $\nk = \sum_{i \in [k] } 1$ indicate the number of units. Let $\nt = \sumi A_i$, $\nc = \sumi (1-A_i)$, $\nkt = \sumik A_i$, and $\nkc = \sumik ( 1 - A_i )$ denote the the numbers of treated units, control units, treated units in stratum $k$, and  control units in stratum $k$, respectively. The proportions of stratum sizes and  treated units in stratum $k$ are denoted as  $p_{n[k]}  = \nk / n$ and $\pik = \nkt / \nk $, respectively. The treatment effect is
$$
\tau = E \{ Y_i(1) - Y_i(0) \} = \sumk \pk \left[  E \{ Y_i(1) \mid B_i = k \} -  E \{ Y_i(0) \mid B_i = k \} \right] = \sumk \pk \tauk,
$$
where $\tauk =  E \{ Y_i(1) \mid B_i = k \} -  E \{ Y_i(0) \mid B_i = k \}$ is the treatment effect in stratum $k$. We aim to improve the estimation efficiency of $\tau$ by using the information present in the high-dimensional covariates $\bx_i$.




Let $\mathcal{L}_2 $ and $ \mathcal{R}_2 $ be sets of random variables with bounded second moments and (strictly) positive stratum-specific variances, respectively.
$$
\mathcal{L}_2 = \{  (V_1,\dots, V_m) :   E( |V_j | ^ 2)  < \infty,  \ j=1,\dots, m \},
$$
$$
 \mathcal{R}_2 = \{  (V_1,\dots, V_m) :  \max_{k=1,\dots,K} \Var \{  V_j | B_j = k   \}   > 0 , \ j=1,\dots,m \},
$$
We assume that the stratum-specific covariance matrix
$$ \Sigma_{[k] \bx \bx } =  E[ \{ \bx_i - E(\bx_i | B_i = k ) \} \{ \bx_i - E(\bx_i | B_i  = k ) \} ^\T \mid B_i = k ] , \quad k=1,\dots,K, $$
is (strictly) positive-definite, and make the following requirements for the  data generating process and treatment assignment mechanism.



\begin{assumption}
\label{assum::Q}
$\{ Y_i(1), Y_i(0), B_i, \bx_i \}_{i=1}^{n} $ are independent and identically distributed  samples from the population distribution of $\{ Y(1), Y(0), B, \bx \}$. Moreover, $\{ Y_i(1), Y_i(0) \} \in \mathcal{L}_2 \cap  \mathcal{R}_2$, and there exists a constant $M$ independent of $n$, such that $ \max_{i=1,\dots,n;\ j = 1,\dots,p} | x_{ij} | \leq M$.
\end{assumption}

\begin{assumption}
\label{assum::A}
The treatment assignment mechanism satisfies the following conditions:
\begin{itemize}
\item[(a)] Conditional on $\{ B_1,\dots, B_n \}$, $ \{ A_1,\dots, A_n \} $ are independent of $ \{ Y_i(1), Y_i(0), \bx_i \}_{i=1}^{n}$ .
\item[(b)] For $k=1,\dots,K$, $\pik$ converges in probability to $ \pi$.
\end{itemize}
\end{assumption}




The above two assumptions are similar to those proposed in \citet{Bugni2019} and \citet{Ma2020Regression}, with the only difference being that $\bx_i$ is high-dimensional and uniformly bounded. In the low-dimensional setting in which $p$ is fixed, the uniformly bounded assumption on $\bx_i$ can be relaxed to $\bx_i \in  \mathcal{L}_2$. In the high-dimensional setting in which $p$ is comparable to, or even larger than, $n$, we make this assumption to weaken the requirements on the approximation errors (see Remark~\ref{remark::bounded} in Section 5). As our main theorems allow the linear model to be arbitrarilly misspecified, the uniformly bounded assumption can be fulfilled by transforming the covariates when they are relatively large. Note that the assumption $\{ Y_i(1), Y_i(0) \} \in  \mathcal{R}_2$ is made only to rule out the degenerate situations of an asymptotically normal distribution. Assumption~\ref{assum::A} is quite general and satisfied by most, if not all, covariate-adaptive randomization methods, such as stratified biased-coin design \citep{Efron1971},  stratified  adaptive biased-coin design \citep{Wei1978}, stratified block randomization \citep{Zelen1974}, Pocock and Simon's minimization \citep{Pocock1975}, and the class of designs proposed by \citet{Hu2012}.
Moreover, the assumptions are trivially satisfied for simple and restricted randomization \citep{Rosenberger2015}.



{\bf Notation}. For a random variable $V$, let us denote its mean and variance as $\mu_{V} = E(V)$ and $\sigma^2_{V} = \Var(V)$, respectively. For random variables $r_i(a)$, such as the potential outcomes $Y_i(a)$, covariates $\bx_i$, or their transformations ($i=1,\dots,n$, $a=0,1$), we add a bar on top and a subscript 1 (or 0) to denote their sample mean under treatment (control); that is,
$
\bar{r}_1 = ( 1 / \nt ) \sumi A_i r_i(1) $ and $ \bar{r}_0 = ( 1 / \nc )  \sumi (1 - A_i ) r_i(0).
$
We add an additional subscript $[k]$ to denote their stratum-specific sample means; that is,
$
 \bar{r}_{[k]1} = ( 1 / \nkt)  \sumik A_i r_i(1) $ and $ \bar{r}_{[k]0} = ( 1 / \nkc ) \sumik (1 - A_i ) r_i(0).
$
The following two quantities are the main components of the asymptotic variance of the treatment effect estimator:
\[
\varsigma^2_{r}(\pi) = \frac{1}{\pi} \sigma^2_{r_i(1) - E\{ r_i(1) \mid B_i \}} + \frac{1}{ 1 - \pi } \sigma^2_{ r_i(0) - E\{ r_i(0) \mid B_i \} },
\]
\[
\varsigma^2_{H r} = \sumk \pk \Big(  \big[ E\{ r_i(1) \mid B_i = k \} - E\{ r_i(1) \} \big]  -  \big[ E\{ r_i(0) \mid B_i = k \} - E\{ r_i(0) \} \big]  \Big)^2.
\]
We denote their sample analog as
\begin{eqnarray}
\hat \varsigma^2_{r}(\pi) &= & \frac{1}{\pi}  \sumk \pnk  \bigg[  \frac{1}{\nkt} \sumik  A_i  \Big\{  \hat r_i(1)  - \frac{1}{\nkt} \sum_{j \in [k]} A_j  \hat r_j(1)  \Big\} ^2  \bigg] \nonumber\\
&& + \frac{1}{1 - \pi} \sumk \pnk  \bigg[  \frac{1}{\nkc} \sumik (1 - A_i ) \Big\{ \hat r_i (0)  -   \frac{1}{\nkc} \sum_{j \in [k]} ( 1 - A_j )  \hat r_j (0)  \Big\}^2   \bigg], \nonumber
\end{eqnarray}
$$
\hat  \varsigma^2_{H r} = \sumk \pnk  \bigg[ \Big\{ \frac{1}{\nkt} \sum_{j \in [k]} A_j  \hat r_j(1)   - \frac{1}{\nt} \sumi A_i \hat r_i (1) \Big\}    -  \Big\{  \frac{1}{\nkc} \sum_{j \in [k]} ( 1 - A_j )  \hat r_j (0) - \frac{1}{\nc}  \sumi ( 1 - A_i ) \hat r_i(0)   \Big\} \bigg]^2,
$$
where $\hat r_i(a)$ is the estimated (or observed) value of $r_i(a)$.  We denote the covariance between two random vectors $R$ and $Q$ as $ \Sigma_{RQ} = E[ \{R - E(R) \} \{ Q - E(Q)  \}^\T ]  $. For a vector $u = (u_1,\dots,u_m)^\T$ and a set $S \subset \{1,\ldots,m\}$, let $|| u ||_1 = \sum_{i=1}^{m} | u_i |$, $|| u ||_2 = ( \sum_{i=1}^{m} u_i^2 )^{1/2}$, and $|| u ||_\infty = \max_{i=1,\dots,m} | u_i |$ denote the $l_1$, $l_2$,  and $l_\infty$ norms, respectively. Let $S^c$ denote the complementary set of $S$, $|S|$ denote the cardinality of $S$, and $u_S = (u_j, j \in S)^\T$ denote the vector of elements of $u$ in $S$.

\section{OLS-adjusted treatment effect estimator}

As the stratum-specific treatment effect $\tauk =  E \{ Y_i(1) \mid B_i = k \} -  E \{ Y_i(0) \mid B_i = k \}$ can be estimated without bias by the difference in the stratum-specific sample means $ \taukhat  = \YkThat - \YkChat $, a  plug-in (and unbiased) estimator for $\tau$ is the stratified difference-in-means:
$$
\taustr = \sumk \pnk \left(   \YkThat - \YkChat \right).
$$
As shown by  \citet{Ma2020Regression}, $\taustr$ can be interpreted as an OLS estimator of the coefficient of $A_i$ in the regression of $Y_i$ on $A_i$, $I_{B_i = k } $ and the interactions $A_i ( I_{B_i=k} - \pnk )$, where $I_{B_i = k}$, $k=1,\dots,K$, are the stratification indicators:
\begin{equation}
\label{reg::str}
Y_i \sim \alpha +  A_i \tau + \sum_{k=1}^{K-1} \alpha_k  I_{ B_i = k } + \sum_{k=1}^{K-1} \nu_k A_i (  I_{ B_i = k } - \pnk ).
\end{equation}
Moreover, $\taustr$ is consistent, asymptotically normal, and  the most efficient estimator among the commonly used regression estimators adjusting (or not adjusting) the stratification indicators $I_{B_i = k }$; see \cite{Ma2020Regression} for a detailed discussion.

\begin{proposition}[\citet{Bugni2019, Ma2020Regression}]
\label{prop::str}
Under  Assumptions~\ref{assum::Q} and \ref{assum::A},
$$
\surd{n} ( \taustr - \tau ) \xrightarrow{d} N \big(0,  \varsigma^2_{ Y}(\pi) + \varsigma^2_{H Y}  \big), \quad  \hat \varsigma^2_{ Y}(\pi) + \hat \varsigma^2_{H Y}  \xrightarrow{P}  \varsigma^2_{ Y}(\pi) + \varsigma^2_{H Y} .
$$
\end{proposition}

The additional covariates $\bx_i$ may contain useful information to improve the estimation efficiency of the treatment effect. Let $\bar{\bx} = (1/n) \sumi \bx_i$ and $\XkT = (1/\nk) \sumik \bx_i$. Under a low-dimensional setting, \citet{Ma2020Regression} proposed a more efficient estimator $\tauols$:
$$
\tauols =  \sumk \pnk \Big[  \Big\{   \YkThat - ( \XkThat -  \XkT )^\T \hat \beta_\ols (1)   \Big\}  - \Big\{  \YkChat - ( \XkChat -  \XkC )^\T \hat \beta_\ols (0)   \Big\}  \Big],
$$
where $\hat \beta_\ols (1)$ and $\hat \beta_\ols (0)$ are the OLS estimators of the $\bx_i$  coefficients when regressing $Y_i$ on $\bx_i$ (with intercept) in the treatment and control groups, respectively. This estimator is equivalent to adding $\bx_i$ and the treatment-by-covariate interactions $A_i ( \bx_i - \bar{\bx} ) $ into regression~\eqref{reg::str}. Although $\tauols$ is $\mathcal{S}$-optimal \citep{Ma2020Regression}, its  efficiency  can be further improved by using stratum-specific adjusted vectors \citep{Ye2020Inference}:
$$
\tauinter =  \sumk \pnk \Big[  \Big\{   \YkThat - ( \XkThat -  \XkT )^\T \hat \beta_{[k] \ols }(1)   \Big\}  - \Big\{  \YkChat - ( \XkChat -  \XkC )^\T \hat \beta_{[k] \ols }(0)   \Big\}  \Big],
$$
where $\hat \beta_{[k] \ols }(1)$ and $\hat \beta_{[k] \ols}(0)$ are the stratum-specific OLS estimators of the coefficients of $\bx_i$ when regressing $Y_i$ on $\bx_i$ (with intercept) in the treatment and control groups  within stratum $k$, respectively.

\section{General theory for regression adjustment}



The OLS estimator does not work in a high-dimensional setting,  due to over-fitting. Thus, the selection of covariates or some form of regularization is necessary for an effective treatment effect estimation. This motivates us to use penalized regression,  such as the Lasso \citep{Tibshirani1996}, to perform the covariate adjustment. In this section, we develop a general theory for a regression-adjusted treatment effect estimator, which will be applied to  two Lasso-adjusted treatment effect estimators in the next sections.

Let $ \hat \beta_{[k]}(1)$ and $ \hat \beta_{[k]}(0)$ be some estimated adjusted vectors (can be the same across strata). Similar to $\tauinter$, the general regression-adjusted treatment effect estimator can be defined as
$$
\taugen =  \sumk \pnk \Big[  \Big\{   \YkThat - ( \XkThat -  \XkT )^\T \hat \beta_{[k]}(1)   \Big\}  - \Big\{  \YkChat - ( \XkChat -  \XkC )^\T \hat \beta_{[k]}(0)   \Big\}  \Big].
$$

We now introduce conditions on $ \hat \beta_{[k]}(1)$ and $ \hat \beta_{[k]}(0)$, which can guarantee the asymptotic normality of $\taugen$.

\begin{assumption}
\label{assum::gen}
For $k=1,\dots,K$, there exist coefficient vectors $\beta_{[k]}(1)$ and $\beta_{[k]}(0)$, such that
$$
|| \hat \beta_{[k]}(a) - \beta_{[k]}(a) ||_{1} = o_P(1), \quad \surd{n} ( \XkThat -  \XkChat )^\T \left\{ \hat \beta_{[k]}(a) - \beta_{[k]}(a)  \right\} = o_P(1), \quad a = 0, 1.
$$
\end{assumption}

\begin{remark}
In a low-dimensional setting, under Assumptions~\ref{assum::Q} and \ref{assum::A}, each element of $\surd{n} ( \XkThat -  \XkChat )$ is asymptotically normal with zero mean and finite variance. Then,  Assumption~\ref{assum::gen} is implied if $ \hat \beta_{[k]}(a) - \beta_{[k]}(a)$ converges (element-wise) to zero in probability. \citet{Ma2020Regression} and \citet{Ye2020Inference} showed that the OLS estimators satisfy this requirement. In contrast, in a high-dimensional setting, by concentration inequality, $\surd{n} ( \XkThat -  \XkChat ) = O_P( \surd{  \log p  })$ (the rigorous proof is given in the Appendix). Then, by H\"older inequality, Assumption~\ref{assum::gen} is implied by $ || \hat \beta_{[k]}(a) - \beta_{[k]}(a) ||_1 = o_P( 1/ \surd{ \log p} ) $. In the next two sections, we will present the conditions under which $ \hat \beta_{[k]}(a) $ obtained from Lasso satisfies this requirement.
\end{remark}

To establish the theoretical properties of $\taugen$, we need to define the following transformed outcomes and projection coefficients. The transformed outcomes $r_{i, \gen}(a)$ and $\r_{i,\gen}(a)$ are defined such that, conditional on $B_i = k$,
$$
r_{i, \gen}(a) =  Y_i(a) - \bx_i ^\T \beta_{[k]}^{*}, \quad \r_{i, \gen}(a) =  Y_i(a) -  \bx_i  ^\T \beta_{[k] }(a), \quad a = 0, 1,
$$
where $\beta_{[k]}^{*} = ( 1 - \pi ) \beta_{[k]}(1) + \pi  \beta_{[k]}(0)$. The  estimated values of $r_{i, \gen}(a)$ are
$$
\hat  r_{i,\gen}(a) = Y_i(a) -   \bx_i   ^\T \hat \beta_{[k]}^{*}, \quad i \in [k], \quad a  = 0, 1,
$$
where $  \hat \beta_{[k]}^{*} = ( 1 - \pik ) \hat \beta_{[k]}(1)  + \pik \hat  \beta_{[k]}(0) $. The stratum-common and stratum-specific projection coefficients can be defined as
\begin{equation}
\label{eqn::tildebeta}
\beta_{\proj} (a) = \argmin_{ \beta } E \big [  Y_i(a) - E \{ Y_i(a) | B_i \} -  \{ \bx_i - E ( \bx_i | B_i )  \} ^\T \beta  \big ]^2,  \nonumber
\end{equation}
\begin{equation}
\label{eqn::tildebeta}
\beta_{[k] \proj}(a) = \argmin_{ \beta } E \Big(  \big [  Y_i(a) - E \{ Y_i(a) | B_i = k \} -  \{ \bx_i - E ( \bx_i | B_i = k )  \} ^\T \beta \big]^2   \mid B_i = k  \Big) .   \nonumber
\end{equation}

\begin{theorem}
\label{thm::gen}
Supposing that $\{  r_{i,\gen}(1), r_{i,\gen}(0) \} \in  \mathcal{R}_2$, $\{  \r_{i,\gen}(1), \r_{i,\gen}(0) \} \in \mathcal{L}_2  $ and Assumptions~\ref{assum::Q}--\ref{assum::gen} hold, 
$$
\surd{n} ( \taugen - \tau ) \xrightarrow{d} N \big(0,  \varsigma^2_{ r_{\gen}}(\pi) + \varsigma^2_{H r_{\gen}}  \big), \quad \hat \varsigma^2_{ r_\gen }(\pi) + \hat  \varsigma^2_{H r_\gen} \xrightarrow{P}  \varsigma^2_{r_{\gen}}(\pi) + \varsigma^2_{H r_{\gen}} .
$$
Furthermore, the asymptotic variance is minimized at $\beta_{[k]}(a) = \beta_{ \proj}(a)$, under the constraint that $\beta_{[k]}(a)$ remain the same across strata, and at $\beta_{[k]}(a) = \beta_{[k] \proj}(a)$, without constraint, $k=1,\dots,K$, $a=0,1$.
\end{theorem}

Theorem~\ref{thm::gen} implies that as long as the estimated adjusted vectors satisfy Assumption~\ref{assum::gen}, the resulting regression-adjusted treatment effect estimator is consistent and asymptotically normal. Moreover, its asymptotic variance has the smallest value when $\beta_{[k]}(a) = \beta_{ \proj}(a)$ for stratum-common adjusted vectors and is minimized at $\beta_{[k]}(a) = \beta_{[k] \proj}(a)$ for stratum-specific adjusted vectors.  \citet{Ma2020Regression} showed that, in a low-dimensional setting, $ \hat \beta_{\ols} (a) - \beta_{\proj} (a) = o_P(1) $, and thus, satisfies Assumption~\ref{assum::gen}. Therefore, $\tauols$ is optimal among the class of stratum-common regression-adjusted estimators
$$  \sumk \pnk \Big[  \Big\{   \YkThat - ( \XkThat -  \XkT )^\T \hat \beta(1)   \Big\}  - \Big\{  \YkChat - ( \XkChat -  \XkC )^\T \hat \beta(0)   \Big\}  \Big] .$$
Moreover, as shown by \citet{Ye2020Inference},  $\hat \beta_{[k]\ols}(a) - \beta_{[k] \proj} (a) = o_P(1)$, and thus, also satisfies Assumption~\ref{assum::gen}. Therefore,  $\tauinter$ is optimal among the class of stratum-specific regression-adjusted estimators
$$   \sumk \pnk \Big[  \Big\{   \YkThat - ( \XkThat -  \XkT )^\T \hat \beta_{[k]}(1)   \Big\}  - \Big\{  \YkChat - ( \XkChat -  \XkC )^\T \hat \beta_{[k]}(0)   \Big\}  \Big] .$$ Theorem~\ref{thm::gen} extends these results to general situations in which penalized (or regularized) regression-adjusted vectors, such as those obtained by the Lasso, can be used to handle high-dimensional covariates. Moreover, the asymptotic variance can be consistently estimated using a non-parametric variance estimator. Thus, we can construct a valid inference for the treatment effect based on $\taugen$ and $ \hat \varsigma^2_{ r_\gen }(\pi) + \hat  \varsigma^2_{H r_\gen} $.  Furthermore, this theorem does not assume a linear model for the true data-generating process; that is, it allows the linear model to be arbitrarily misspecified. Based on this theorem, in the next two sections, we will study two Lasso-adjusted treatment effect estimators by using the stratum-common and stratum-specific Lasso-adjusted vectors, respectively.

\section{Stratum-common Lasso-adjusted treatment effect estimator}

Similar to $\tauols$, we define the stratum-common Lasso-adjusted treatment effect estimator by replacing the OLS-adjusted vectors with the Lasso-adjusted vectors:
\begin{equation}
\label{eqn:Lasso1}
\betaLassoT = \argmin_{ \beta } \frac{1}{2 \nt} \sumk \sumik A_i \big\{ Y_i - \YkThat -  (\bx_i - \XkThat )^\T \beta \big\}^2 + \lambda_1 || \beta ||_1, \nonumber
\end{equation}
\begin{equation}
\label{eqn:Lasso0}
\betaLassoC = \argmin_{ \beta } \frac{1}{2 \nc} \sumk \sumik ( 1 - A_i ) \big\{ Y_i - \YkChat -  (\bx_i - \XkChat )^\T \beta \big\}^2 + \lambda_0 || \beta ||_1. \nonumber
\end{equation}
The stratum-common Lasso-adjusted  treatment effect estimator can be defined as
$$
\tauLasso = \sumk \pnk  \Big[  \big\{   \YkThat - ( \XkThat -  \XkT )^\T \betaLassoT   \Big\}  - \big\{  \YkChat - ( \XkChat -  \XkC )^\T  \betaLassoC   \Big\}  \Big].
$$
Clearly, $\tauLasso$ belongs to the class of regression-adjusted treatment effect estimators of the form
$$
\sumk \pnk  \Big[  \big\{   \YkThat - ( \XkThat -  \XkT )^\T  \hat \beta(1)  \Big\}  - \big\{  \YkChat - ( \XkChat -  \XkC )^\T  \hat \beta(0)   \Big\}  \Big].
$$

To investigate the asymptotic properties of $\tauLasso$, we need to outline the conditions under which the Lasso-adjusted vectors $\betaLassoT$ and $\betaLassoC$ satisfy Assumption~\ref{assum::gen}. For this purpose, as  not all covariates are relevant to the potential outcomes  in many high-dimensional problems, it is common and reasonable to assume that the projection coefficients $\beta_{\proj}(1)$ and $\beta_{\proj}(0)$ are sparse. We denote the set of relevant covariates as $S = \{ j \in \{1,\dots,p\} : \beta_{j,\proj} (1) \neq 0 \ \textnormal{or} \ \beta_{j, \proj}(0) \neq 0 \}$ and $s = | S | $ as the total number of relevant covariates. Then, the transformed outcomes are defined as
$$
\r_i(a) = Y_i(a) -  \bx_i  ^\T \beta_{\proj} (a), \quad a = 0, 1.
$$

To establish  the asymptotic normality of $\tauLasso $, we need to study the $l_1$ convergence rates of the Lasso-adjusted vectors $\betaLassoT$ and $\betaLassoC$ under covariate-adaptive randomization, allowing the linear model to be arbitrarily misspecified. For this purpose, we make the following assumptions.

\begin{assumption}
\label{assum::sub-Gaussian}
The stratum-specific covariance matrix $\Sigma_{[k] \bx  \bx} $ satisfies the restricted eigenvalue condition; that is, there exists a constant $c_{RE}$ independent of $n$, such that, for all $\bh \in \mathcal{C} = \{  \bh \in R^p :  || \bh_{S^c} ||_1 \leq  3 || \bh_{S} ||_1 \}$, it holds that
$
\bh ^\T \Sigma_{[k] \bx  \bx} \bh \geq c_{RE} || \bh ||_2^2, \ k = 1,\dots,K
$
\end{assumption}

\begin{assumption}
\label{assum::lambda}
There exist constants $c_{\lambda} > 0$ (defined in the proof), $M > 1$ and sequence $M_n \rightarrow \infty$, such that the tuning parameters $\lambda_1$ and $\lambda_0$ belong to the following interval:
$$
\left[ 4 \sumk  \bigg( \frac{c_{\lambda} \pk M_n  }{ \pi} \bigg)^{1/2}  \cdot \bigg (  \frac{\log p }{ n }   \bigg)^{1/2}  , \  4M \sumk  \bigg( \frac{c_{\lambda} \pk M_n }{ \pi} \bigg)^{1/2} \cdot \bigg (  \frac{\log p }{ n }   \bigg)^{1/2} \right].
$$
\end{assumption}

\begin{assumption}
\label{assum::sparsity}
Suppose that $ M_n s^2 ( \log p )^2 / n  \rightarrow 0$.
\end{assumption}

\begin{remark}
The sample-version restricted eigenvalue condition (or  its variants) is a typical assumption for obtaining the $l_1$ convergence rate of  Lasso in high-dimensional sparse linear regression models \citep[e.g.,][]{Zhang2008, Huang2008, Meinshausen2009, Negahban2009}. As we consider random covariates, we require the population-version restricted eigenvalue condition in Assumption~\ref{assum::sub-Gaussian}, and we will show that it implies the sample-version restricted eigenvalue condition with probability tending to one under covariate-adaptive randomization. The sequence $M_n$ in Assumption~\ref{assum::lambda} can tend to infinity very slowly; for example, $M_n = \log \log n$. Thus, the main requirement on the tuning parameters is that they have the order of $\{(\log p)/ n\}^{1/2}$ (expect for the factor $M_n^{1/2}$), which is typically assumed for Lasso in high-dimensional sparse linear regression models. We will explain later why we need the extra factor $M_n^{1/2}$. Assumption~\ref{assum::sparsity} (except for the factor $M_n$), typically required in inference using the de-biased Lasso \citep{Zhang2014debiased}, is stronger by a factor of $ \log p $ than the condition for obtaining the $l_1$ consistency of  Lasso.

\end{remark}


\begin{theorem}
\label{thm::betaLasso}
Suppose that $\{  \r_i(1), \r_i(0) \} \in \mathcal{L}_2 $ and Assumptions~\ref{assum::Q}--\ref{assum::A} and \ref{assum::sub-Gaussian}--\ref{assum::sparsity} hold, then
$$
|| \hat \beta_{\Lasso}(a) - \beta_{\proj}(a) ||_1 = O_P\left\{  s  \bigg( \frac{ M_n \log p}{n} \bigg)^{1/2} \right\}, \quad a= 0, 1.
$$
\end{theorem}


Theorem~\ref{thm::betaLasso} establishes  the $l_1$ convergence rate of  Lasso under covariate-adaptive randomization. The same convergence rate (except for the factor $M_n^{1/2}$) is obtained under a high-dimensional spare linear regression model with fixed covariates $\bx_i$ and independent and identically distributed Gaussian (or sub-Gaussian) random errors; see, for example, \citet{Zhang2008}, \citet{Meinshausen2009}, and \citet{Negahban2009}. \citet{Bloniarz2016} established a similar convergence rate under a finite-population framework and simple randomization. Theorem~\ref{thm::betaLasso} extends these results to random covariates and dependent observations under general covariate-adaptive randomization, allowing the linear regression model to be arbitrarily misspecified.

\begin{remark}\label{remark::bounded}
The uniformly bounded condition on the covariates $\bx_i$ can be relaxed to the sub-Gaussian condition if we assume a stronger condition on the transformed outcomes $ \r_i(a)$, for example, assuming that both $ \r_i(a)$ and $\bx_i \r_i(a)$ are sub-Gaussian random variables. In this case,  the extra factor $M_n^{1/2}$ will disappear.
\end{remark}


Based on Theorem~\ref{thm::betaLasso}, we can establish the asymptotic normality of $\tauLasso$ and the consistency of its variance estimator. For $a = 0,1$, define the transformed outcomes and their estimated values by
$$ r_i(a) = Y_i(a) - \bx_i ^\T \beta_{ \proj} ^{*}, \quad \hat  r_i(a) = Y_i(a)  -   \bx_i  ^\T \hat \beta_{\Lasso}^{*}, \quad i \in [k] , $$
where
$ \beta_{\proj}^ {*} = ( 1 - \pi ) \beta_{\proj}(1) + \pi \beta_{\proj}(0)$ and $  \hat \beta_{\Lasso}^{*} = ( 1 - \pik ) \betaLassoT + \pik  \betaLassoC. $


\begin{theorem}
\label{thm::Lasso}
Suppose that $\{ r_i(1), r_i(0) \} \in \mathcal{R}_2$, $ \{  \r_i(1), \r_i(0) \} \in \mathcal{L}_2  $, and Assumptions~\ref{assum::Q}--\ref{assum::A} and \ref{assum::sub-Gaussian}--\ref{assum::sparsity} hold, then
$$
\surd{n} ( \tauLasso - \tau ) \xrightarrow{d} N \big(0,  \varsigma^2_{ r}(\pi) + \varsigma^2_{H r}  \big), \quad \hat \varsigma^2_{ r}(\pi) + \hat \varsigma^2_{H r} \xrightarrow{P} \varsigma^2_{ r}(\pi) + \varsigma^2_{H r}.
$$
Furthermore, the difference between the asymptotic variances of $\tauLasso$ and $\taustr$ is
$$
\Delta =   -  \frac{1}{\pi ( 1 - \pi ) }  ( \beta_{\proj}^ {*})^\T \Sigma_{\tilde \bx \tilde \bx} ( \beta_{\proj}^ {*} ) \leq 0,
$$
where $\tilde \bx = \bx - E( \bx \mid B )$.
\end{theorem}

Theorem~\ref{thm::Lasso} shows that, under appropriate conditions, the stratum-common Lasso-adjusted treatment effect estimator $\tauLasso$ performs as if the true projection coefficients $\beta_{\proj}(1)$ and $\beta_{\proj}(0)$ are known. Moreover, $\tauLasso$ improves, or at least does not degrade, the precision when compared with the stratified difference-in-means estimator ($\taustr$)  without adjusting for the additional covariates $\bx_i$. Combined with Theorem~\ref{thm::gen}, it is optimal among the stratum-common regression-adjusted treatment effect estimators. Moreover, the asymptotic variance of $\tauLasso$ can be consistently estimated using a non-parametric variance estimator. Therefore, based on $\tauLasso$ and $  \hat \varsigma^2_{ r}(\pi) + \hat \varsigma^2_{H r} $, we can make a robust and efficient inference for the treatment effect under covariate-adaptive randomization with high-dimensional covariates.



\begin{remark}\label{rem::adj1}
In a finite sample, the variance estimator $ \hat \varsigma^2_{r}(\pi) + \hat \varsigma^2_{H r} $ may under-estimate the asymptotic variance. This drawback can be partly solved by adjusting for the degrees of freedom of the Lasso-adjusted vectors, following the ideas presented in \citet{Bloniarz2016}. That is, letting $\hat s(a) = |   \{  j \in 1,\dots,p: \ \hat \beta_{j,\Lasso}(a) \neq 0  \} |$ denote the number of covariates selected by Lasso,  $  \hat \varsigma^2_{r}(\pi)  $ can be replaced by
\begin{eqnarray}
\hat \varsigma^2_{r, \adj}(\pi) & = &    \frac{n}{n - \hat s(1)  - 1} \cdot  \frac{1}{\pi} \sumk \pnk  \bigg[  \frac{ 1 }{\nkt } \sumik  A_i  \Big\{  \hat r_i(1)  - \frac{1}{\nkt} \sum_{j \in [k]} A_j  \hat r_j(1)  \Big\} ^2  \bigg]  +\nonumber\\
&&    \frac{n}{n - \hat s(0)  - 1} \cdot \frac{1}{1 - \pi} \sumk  \pnk  \bigg[  \frac{ 1  }{\nkc } \sumik (1 - A_i )  \Big\{ \hat r_i (0)  -   \frac{1}{\nkc} \sum_{j \in [k]} ( 1 - A_j )  \hat r_j (0)  \Big\}^2   \bigg]. \nonumber
\end{eqnarray}
\end{remark}

\section{Stratum-specific Lasso-adjusted treatment effect estimator}

The stratum-common Lasso-adjusted treatment effect estimator $\tauLasso$ uses the same Lasso-adjusted vectors $\betaLassoT$ and $\betaLassoC$ for different strata. As shown in Theorem~\ref{thm::gen}, the estimation efficiency can be further improved by using the stratum-specific Lasso-adjusted vectors, as least asymptotically. More specifically, for $k=1,\dots,K$,  the stratum-specific Lasso-adjusted vectors can be defined as

\begin{equation}
\label{eqn:Lasso1}
\betaLassoTs = \argmin_{ \beta } \frac{1}{2 \nkt} \sumik A_i \big\{ Y_i - \YkThat -  (\bx_i - \XkThat )^\T \beta \big\}^2 + \lambda_{[k]1} || \beta ||_1, \nonumber
\end{equation}
\begin{equation}
\label{eqn:Lasso0}
\betaLassoCs = \argmin_{ \beta } \frac{1}{2 \nkc} \sumik ( 1 - A_i ) \big\{ Y_i - \YkChat -  (\bx_i - \XkChat )^\T \beta \big\}^2 + \lambda_{[k]0} || \beta ||_1. \nonumber
\end{equation}
The stratum-specific Lasso-adjusted treatment effect estimator can be defined as
$$
\tauLassos = \sumk \pnk  \Big[  \big\{   \YkThat - ( \XkThat -  \XkT )^\T \betaLassoTs   \Big\}  - \big\{  \YkChat - ( \XkChat -  \XkC )^\T  \betaLassoCs   \Big\}  \Big].
$$

To investigate the asymptotic properties of $\tauLassos$, we define the transformed outcomes $\eta_{i}(a)$, such that conditional on $B_i = k$,
$$
\eta_{i}(a) =  Y_i(a) -    \bx_i  ^\T \beta_{[k]\proj} (a), \quad a = 0, 1.
$$
In fact, $\eta_{i}(a)  - E \{  \eta_{i}(a)  \mid B_i = k \}$ is the  error of projecting the potential outcomes onto the space spanned by the (relevant) covariates within stratum $k$. Let $S_{[k]} = \{ j \in \{1,\dots,p\} : \beta_{j,[k] \proj} (1) \neq 0 \ \textnormal{or} \ \beta_{j, [k] \proj}(0) \neq 0 \} $ and let $s_{[k]} = | S_{[k]} | $ be the total number of relevant covariates in stratum $k$. We require the following conditions within each stratum to obtain the $l_1$ convergence rates of the stratum-specific Lasso-adjusted vectors $\betaLassoTs$ and $\betaLassoCs$, which ensure  that those adjusted vectors satisfy Assumption~\ref{assum::gen} with respect to $\beta_{[k]\proj} (1)$ and $\beta_{[k]\proj} (0) $.


\begin{assumption}
\label{assum::lambda2}
There exist constants $c_{[k] \lambda} > 0$, $M > 1$, and sequence $M_{[k]n} \rightarrow \infty$, such that the tuning parameters $\lambda_{[k]1}$ and $\lambda_{[k]0}$ belong to the following interval:
$$
\left[ 4 \bigg( \frac{c_{[k] \lambda} \pk M_{[k]n} }{ \pi} \bigg)^{1/2} \cdot \bigg (  \frac{\log p }{ n }   \bigg)^{1/2}  , \  4M \bigg( \frac{c_{[k] \lambda} \pk M_{[k]n} }{ \pi} \bigg)^{1/2} \bigg (  \frac{\log p }{ n }   \bigg)^{1/2} \right].
$$
\end{assumption}

\begin{assumption}
\label{assum::sparsity2}
Suppose that $ M_{[k]n}  s^2_{[k]} ( \log p )^2 / n  \rightarrow 0$, $k=1,\dots,K$.
\end{assumption}

Assumptions~\ref{assum::lambda2} and \ref{assum::sparsity2} are the stratum-specific analogs of Assumptions~\ref{assum::lambda} and \ref{assum::sparsity}, respectively. Using these assumptions, we can apply Theorem~\ref{thm::betaLasso} (with $K=1$) to each stratum $k$, and obtain the following corollary on the $l_1$ convergence rates of the stratum-specific Lasso-adjusted vectors, which are crucial for proving the asymptotic normality of $\tauLassos$.

\begin{corollary}
\label{cor::betaLasso}
Suppose that $\{  \eta_{i}(1), \eta_{i}(0) \} \in \mathcal{L}_2 $ and Assumptions~\ref{assum::Q}-- \ref{assum::A},  \ref{assum::sub-Gaussian}, and \ref{assum::lambda2}--\ref{assum::sparsity2} hold, then
$$
|| \hat \beta_{[k]\Lasso}(a)  - \beta_{[k] \proj}(a) ||_1 = O_P\left\{  s_{[k]} \bigg( \frac{ M_{[k]n} \log p}{n} \bigg)^{1/2} \right\}, \quad k=1,\dots,K, \quad a = 0, 1.
$$
\end{corollary}

Now, we can obtain the asymptotic properties of $\tauLassos$, which depend on the following transformed outcomes $u_{i}(a)$ and their estimated values $\hat  u_{i}(a)$:  conditional on $B_i = k$,
$$
u_{i}(a) = Y_i(a) - \bx_i ^\T \beta_{[k] \proj}^{*}, \quad \hat  u_{i}(a) = Y_i(a)  -   \bx_i  ^\T \hat \beta_{[k] \Lasso}^{*}, \quad a=0, 1,
$$
where $\beta_{[k] \proj}^{*} = ( 1 - \pi ) \beta_{[k] \proj}(1) +  \pi  \beta_{[k] \proj }(0)$ and  $\hat \beta_{[k] \Lasso}^{*} = ( 1 -  \pik )  \hat \beta_{[k] \Lasso}(1) +  \pik  \hat \beta_{[k] \Lasso }(0)$.

\begin{theorem}
\label{thm::Lasso2}
Suppose that $\{  u_{i}(1), u_{i}(0)  \} \in  \mathcal{R}_2 $, $\{  \eta_{i}(1), \eta_{i}(0) \} \in \mathcal{L}_2 $, and Assumptions~\ref{assum::Q}-- \ref{assum::A},  \ref{assum::sub-Gaussian}, and \ref{assum::lambda2}--\ref{assum::sparsity2} hold, then
$$
\surd{n} ( \tauLassos - \tau ) \xrightarrow{d} N \big(0,  \varsigma^2_{ u}(\pi) + \varsigma^2_{H u}  \big), \quad \hat \varsigma^2_{ u}(\pi) + \hat \varsigma^2_{H u} \xrightarrow{P} \varsigma^2_{u}(\pi) + \varsigma^2_{H u}.
$$
Furthermore, the difference between the asymptotic variances of $\tauLassos$ and $\tauLasso$ is
$$
\Delta^* =  -  \frac{1}{\pi ( 1 - \pi ) }  \left\{ \sumk \pk  (\beta_{[k]\proj}^{*} ) ^\T \Sigma_{[k] \bx \bx} (  \beta_{[k] \proj}^{*} )  - ( \beta_{\proj}^{*} ) ^\T \Sigma_{ \tilde \bx \tilde \bx} ( \beta_{\proj}^{*} ) \right\} \leq 0.
$$
\end{theorem}

Theorem~\ref{thm::Lasso2} implies that the stratum-specific Lasso-adjusted treatment effect estimator $\tauLassos$ is consistent and asymptotically normal, and its asymptotic variance can be consistently estimated using a non-parametric variance estimator. Moreover, if the strata are homogeneous in the sense that $\beta_{[k]\proj}(a) = \beta_{\proj}(a)$ for $k=1,\dots,K$ and $a=0,1$, then $ \Delta^*=0$; that is, $\tauLassos$ is asymptotically equivalent to $\tauLasso$. Generally, $\tauLassos$ is more efficient than $\tauLasso$ (and $\taustr$), at least asymptotically. In fact, it is the optimal estimator among the class of estimators of the  form (see Theorem~\ref{thm::gen})
$$   \sumk \pnk \Big[  \Big\{   \YkThat - ( \XkThat -  \XkT )^\T \hat \beta_{[k]}(1)   \Big\}  - \Big\{  \YkChat - ( \XkChat -  \XkC )^\T \hat \beta_{[k]}(0)   \Big\}  \Big] .$$
Based on Theorem~\ref{thm::Lasso2}, we can conduct valid and more efficient inference for the treatment effect under covariate-adaptive randomization with high-dimensional covariates.

\begin{remark}\label{rem::adj2}
In a finite sample, the variance estimator $ \hat \varsigma^2_{ u}(\pi) + \hat \varsigma^2_{H u} $ can be improved by adjusting for the degrees of freedom of the stratum-specific Lasso-adjusted vectors,  following the ideas presented in \citet{Bloniarz2016}. That is, letting $\hat s_{[k]}(a) = |   \{  j \in 1,\dots,p: \ \hat \beta_{j,[k]\Lasso} (a) \neq 0  \} |$ denote the number of covariates selected by  Lasso in stratum $k$,  $\hat \varsigma^2_{u}(\pi)$ can be replaced by
\begin{eqnarray}
\hat \varsigma^2_{u, \adj}(\pi) &= & \frac{1}{\pi}  \sumk   \bigg[  \frac{ \pnk }{ \nkt - \hat s_{[k]}(1) - 1 } \sumik  A_i  \Big\{  \hat u_i(1)  - \frac{1}{\nkt} \sum_{j \in [k]} A_j  \hat u_j(1)  \Big\} ^2  \bigg] \nonumber\\
&& + \frac{1}{1 - \pi} \sumk  \bigg[  \frac{\pnk }{ \nkc - \hat s_{[k]}(0) - 1 } \sumik (1 - A_i ) \Big\{ \hat u_i (0)  -   \frac{1}{\nkc} \sum_{j \in [k]} ( 1 - A_j )  \hat u_j (0)  \Big\}^2   \bigg]. \nonumber
\end{eqnarray}
\end{remark}






\section{Simulation study}\label{SimStudy}

In this section, we examine the empirical performance of  five regression-adjusted estimators of the treatment effect through a simulation study. For $a \in \{0,1\}, i = 1, \dots, n$, the potential outcomes are generated according to the equation
$$
Y_i(a) = \mu_a + g_a(X_i) + \sigma_a(X_i)\varepsilon_{a,i},
$$
where $X_i$, $g_a(X_i)$, and $ \sigma_a(X_i)$ are specified below for three different models. In each model, $(X_i, \varepsilon_{0,i}, \varepsilon_{1,i})$ are independent and identically distributed (i.i.d.), and both $\varepsilon_{0,i}$ and $\varepsilon_{1,i}$ follow the standard normal distribution. In addition, $X_i$'s are used as covariates for the OLS-adjusted estimators, and we generate additional covariates for the Lasso-adjusted estimators.

Model 1: $X_i$ is a two-dimensional vector,

$$\begin{aligned}
g_0(X_i) &= \beta_1 X_{i1} + \beta_2 X_{i1}X_{i2}, \\
g_1(X_i) &= \beta_1 X_{i1} + \beta_2 X_{i1}X_{i2},
\end{aligned}$$
 where $X_{i1}$ takes values in $\{1, 2\}$ with probabilities $0.4$ and $0.6$, $X_{i2} \sim \textup{Unif}[-2,2]$, and they are independent of each other. We set $\sigma_0(X_i) = 3, \ \sigma_1(X_i) = 5$, and $ \beta = (10, 20)^T$. $X_{i1}$ is used for randomization, resulting in two strata. The additional covariates are independent of $X_{i1}$ and $X_{i2}$, and they follow a multivariate normal distribution with zero mean and identity covariance matrix.

Model 2: $X_i$ is a four-dimensional vector,

$$\begin{aligned}
g_0(X_i) &= \sum\limits_{j=1}^4 \beta_j X_{ij}, \\
g_1(X_i) &= \beta_1 \log(X_{i1})X_{i4},
\end{aligned}$$
 where $X_{i1} \sim \textup{Beta}(3,4)$, $X_{i2} \sim \textup{Unif}[-2,2]$, $X_{i3} = X_{i1} X_{i2}$, $X_{i4}$ takes values in $\{3, 5\}$ with probabilities $0.6$ and $0.4$, and $X_{i1}$, $X_{i2}$, and $ X_{i4}$ are independent of each other. We set $\sigma_0(X_i) = X_{i3S}, \ \sigma_1(X_i) = 2X_{i2S}$, and $ \beta = (15, 7, 5, 6)^T$, where $X_{i2S}$ is a stratified variable of $X_{i2}$; if $X_{i2} >1, X_{i2S} = 2$, and otherwise, $X_{i2S} = 1$;  $X_{i3S}$ is a stratified variable of $X_{i3}$; if $X_{i3} >0, X_{i3S} = 2$, and otherwise, $X_{i3S} = 1$. $X_{i2S}$ and $X_{i4}$ are used for randomization, resulting in four strata. The additional covariates are independent of the $X_i$'s, and they follow a multivariate normal distribution with zero mean and the covariance matrix being a symmetric Toeplitz matrix whose first row is a geometric sequence with initial value $1$ and common ratio 0.5.

Model 3:
$X_i$ is a five-dimensional vector, $$g_0(X_i) = g_1(X_i) = \sum\limits_{j=1}^5 \beta_j X_{ij},$$ where $X_{i1} \sim \textup{Beta}(2,2)$, $X_{i2}$ takes values in $\{1, 2, 3, 4\}$ with equal probability, $X_{i3} \sim \textup{Unif}[-2,2]$, $X_{i4}$ takes values in $\{1, 2, 3\}$ with probabilities $0.3$, $0.6$, and  $0.1$, and $X_{i5} \sim \mathcal{N}(0,1)$; all of them are  independent of each other. We set $\sigma_0(X_i) = 1, \ \sigma_1(X_i) = 3$, and $ \beta = (2, 8, 10, 3, 6)^T$. $X_{i2}$ and $X_{i4}$ are used for randomization, resulting in 12 strata. The additional covariates are independent of the $X_i$'s, and they follow a multivariate normal distribution with zero mean and identity covariance matrix.

Here, we present the simulation results of five regression-adjusted treatment effect estimators under simple randomization, stratified block randomization, and Pocock and Simon's minimization for equal allocation. We consider two different sample sizes $n = 200$ and $n = 500$. The dimension of the covariate $p$ used for Lasso is $100$ in both cases, and the block size used in stratified block randomization is $6$. The biased-coin probability $0.75$ and equal weights are used in Pocock and Simon's minimization. In model $1$, Pocock and Simon's minimization is reduced to a stratified biased-coin design, because there is only one stratum for randomization. The bias, standard deviation (SD) of the treatment effect estimators, standard error (SE) estimators, and the empirical coverage probabilities (CP) are computed using $5,000$ replications. We consider both unadjusted and adjusted variance estimators. Please refer to remarks $4$ and $5$ for details regarding the adjustments for the variances of the Lasso-adjusted treatment effect estimators. The adjustment for the variances of the OLS-adjusted treatment effect estimators is performed similarly by replacing the number of covariates selected by Lasso by the actual number of covariates. Similar simulation results for an unequal allocation ($\pi = 2/3$) can be found in the Appendix.

\begin{table}[H]
	\centering
	\caption{Simulated bias, standard deviation, standard error, and coverage probability for different estimators and randomization methods under equal allocation ($\pi=1/2$) and sample size $n = 200$.}\label{equalsim}
	\vskip 5mm
	\begin{threeparttable}
		\setlength{\tabcolsep}{2pt}
\resizebox{\textwidth}{50mm}{
		\begin{tabular}{llcccccccccccccccccccccccc}
		\cline{1-20}
		&  & \multicolumn{6}{c}{Simple Randomization} & \multicolumn{6}{c}{Stratified Block Randomization} & \multicolumn{6}{c}{Minimization} \\ \cline{3-20}
		&
		&
		\multicolumn{1}{c}{Bias} &
		\multicolumn{1}{c}{SD} &
		\multicolumn{2}{c}{SE} &
        \multicolumn{2}{c}{CP} &
		\multicolumn{1}{c}{Bias} &
		\multicolumn{1}{c}{SD} &
		\multicolumn{2}{c}{SE} &
        \multicolumn{2}{c}{CP} &
		\multicolumn{1}{c}{Bias} &
		\multicolumn{1}{c}{SD} &
		\multicolumn{2}{c}{SE} &
        \multicolumn{2}{c}{CP}  \\ \cline{5-6} \cline{7-8} \cline{11-12} \cline{13-14} \cline{17-18} \cline{19-20}
		Model &
		Estimator &
		\multicolumn{1}{c}{} &
		\multicolumn{1}{c}{} &
		\multicolumn{1}{c}{unadj} &
		\multicolumn{1}{c}{adj} &
		\multicolumn{1}{c}{unadj} &
		\multicolumn{1}{c}{adj} &
		\multicolumn{1}{c}{} &
		\multicolumn{1}{c}{} &
		\multicolumn{1}{c}{unadj} &
		\multicolumn{1}{c}{adj} &
		\multicolumn{1}{c}{unadj} &
		\multicolumn{1}{c}{adj} &
		\multicolumn{1}{c}{} &
		\multicolumn{1}{c}{} &
		\multicolumn{1}{c}{unadj} &
		\multicolumn{1}{c}{adj} &
		\multicolumn{1}{c}{unadj} &
		\multicolumn{1}{c}{adj}   \\ \hline
  1&$\taustr$ & 0.08 & 5.48 & 5.47 & - & 0.94 & - & 0.05 & 5.56 & 5.44 & - & 0.95 & - & -0.08 & 5.46 & 5.45 & - & 0.95 & - \\
  &$\tauols$ & 0.01 & 1.71 & 1.68 & 1.70 & 0.95 & 0.95 & 0.01 & 1.71 & 1.68 & 1.70 & 0.94 & 0.95 & 0.00 & 1.72 & 1.68 & 1.70 & 0.94 & 0.94 \\
  &$\tauinter$ & 0.00 & 0.59 & 0.58 & 0.59 & 0.94 & 0.95 & 0.01 & 0.58 & 0.57 & 0.59 & 0.95 & 0.95 & -0.01 & 0.59 & 0.57 & 0.59 & 0.94 & 0.95 \\
  &$\tauLasso$ & 0.02 & 1.85 & 1.82 & 1.85 & 0.94 & 0.95 & 0.01 & 1.86 & 1.82 & 1.84 & 0.94 & 0.94 & -0.01 & 1.85 & 1.82 & 1.84 & 0.95 & 0.95 \\
  &$\tauLassos$ & 0.01 & 0.69 & 0.66 & 0.69 & 0.94 & 0.95 & 0.02 & 0.69 & 0.66 & 0.68 & 0.94 & 0.95 & -0.02 & 0.69 & 0.66 & 0.68 & 0.94 & 0.95 \\
  2&$\taustr$ & -0.05 & 3.40 & 3.32 & - & 0.94 & - & -0.01 & 3.42 & 3.31 & - & 0.94 & - & -0.04 & 3.38 & 3.31 & - & 0.94 & - \\
  &$\tauols$ & 0.06 & 2.62 & 2.56 & 2.59 & 0.94 & 0.94 & 0.06 & 2.63 & 2.56 & 2.59 & 0.94 & 0.95 & 0.08 & 2.58 & 2.56 & 2.60 & 0.94 & 0.95 \\
  &$\tauinter$ & 0.43 & 2.64 & 2.57 & 2.75 & 0.94 & 0.95 & 0.40 & 2.78 & 2.50 & 2.68 & 0.93 & 0.95 & 0.44 & 2.56 & 2.52 & 2.70 & 0.94 & 0.95 \\
  &$\tauLasso$ & -0.02 & 2.79 & 2.58 & 2.88 & 0.93 & 0.95 & 0.00 & 2.80 & 2.58 & 2.88 & 0.93 & 0.96 & 0.00 & 2.75 & 2.58 & 2.88 & 0.93 & 0.96 \\
  &$\tauLassos$ & -0.05 & 3.37 & 3.25 & 3.35 & 0.94 & 0.94 & -0.02 & 3.39 & 3.25 & 3.34 & 0.94 & 0.95 & -0.04 & 3.35 & 3.25 & 3.34 & 0.94 & 0.94 \\
  3&$\taustr$ & 0.01 & 1.91 & 1.81 & - & 0.94 & - & -0.03 & 1.88 & 1.80 & - & 0.93 & - & -0.05 & 1.87 & 1.80 & - & 0.94 & - \\
  &$\tauols$ & 0.00 & 0.32 & 0.31 & 0.32 & 0.94 & 0.94 & -0.01 & 0.32 & 0.30 & 0.31 & 0.93 & 0.94 & 0.00 & 0.32 & 0.30 & 0.31 & 0.93 & 0.94 \\
  &$\tauinter$ & 0.00 & 3.50 & 0.73 & 1.18 & 0.84 & 0.97 & -0.01 & 1.73 & 0.58 & 0.96 & 0.85 & 0.97 & 0.04 & 4.56 & 0.68 & 1.11 & 0.84 & 0.96 \\
  &$\tauLasso$ & 0.00 & 0.37 & 0.34 & 0.36 & 0.93 & 0.94 & -0.01 & 0.36 & 0.33 & 0.35 & 0.93 & 0.94 & 0.00 & 0.36 & 0.33 & 0.35 & 0.93 & 0.94 \\
  &$\tauLassos$ & 0.02 & 1.74 & 1.55 & 1.77 & 0.92 & 0.95 & -0.03 & 1.71 & 1.53 & 1.73 & 0.92 & 0.95 & -0.04 & 1.70 & 1.54 & 1.75 & 0.93 & 0.96 \\
  \hline
		\end{tabular}}
\begin{tablenotes}
\item Note: SD, standard deviation; SE, standard error; CP, coverage probability; \\
           \hspace{2cm} \hphantom{Note:} unadj, unadjusted variance estimator; adj, adjusted variance estimator; \\
           \hspace{2cm} \hphantom{Note:} -, not available.
\end{tablenotes}
\end{threeparttable}
\end{table}

\begin{table}[H]
	\centering
	\caption{Simulated bias, standard deviation, standard error, and coverage probability for different estimators and randomization methods under equal allocation ($\pi=1/2$) and sample size $n = 500$.}\label{equalsim2}
	\vskip 5mm
	\begin{threeparttable}
		\setlength{\tabcolsep}{2pt}
\resizebox{\textwidth}{50mm}{
		\begin{tabular}{llcccccccccccccccccccccccc}
		\cline{1-20}
		&  & \multicolumn{6}{c}{Simple Randomization} & \multicolumn{6}{c}{Stratified Block Randomization} & \multicolumn{6}{c}{Minimization} \\ \cline{3-20}
		&
		&
		\multicolumn{1}{c}{Bias} &
		\multicolumn{1}{c}{SD} &
		\multicolumn{2}{c}{SE} &
        \multicolumn{2}{c}{CP} &
		\multicolumn{1}{c}{Bias} &
		\multicolumn{1}{c}{SD} &
		\multicolumn{2}{c}{SE} &
        \multicolumn{2}{c}{CP} &
		\multicolumn{1}{c}{Bias} &
		\multicolumn{1}{c}{SD} &
		\multicolumn{2}{c}{SE} &
        \multicolumn{2}{c}{CP}  \\ \cline{5-6} \cline{7-8} \cline{11-12} \cline{13-14} \cline{17-18} \cline{19-20}
		Model &
		Estimator &
		\multicolumn{1}{c}{} &
		\multicolumn{1}{c}{} &
		\multicolumn{1}{c}{unadj} &
		\multicolumn{1}{c}{adj} &
		\multicolumn{1}{c}{unadj} &
		\multicolumn{1}{c}{adj} &
		\multicolumn{1}{c}{} &
		\multicolumn{1}{c}{} &
		\multicolumn{1}{c}{unadj} &
		\multicolumn{1}{c}{adj} &
		\multicolumn{1}{c}{unadj} &
		\multicolumn{1}{c}{adj} &
		\multicolumn{1}{c}{} &
		\multicolumn{1}{c}{} &
		\multicolumn{1}{c}{unadj} &
		\multicolumn{1}{c}{adj} &
		\multicolumn{1}{c}{unadj} &
		\multicolumn{1}{c}{adj}   \\ \hline
  1&$\taustr$ & 0.06 & 3.46 & 3.47 & - & 0.95 & - & -0.09 & 3.47 & 3.47 & - & 0.94 & - & -0.05 & 3.41 & 3.47 & - & 0.95 & - \\
  &$\tauols$ & -0.02 & 1.09 & 1.07 & 1.08 & 0.94 & 0.94 & -0.03 & 1.07 & 1.07 & 1.08 & 0.95 & 0.95 & -0.01 & 1.07 & 1.07 & 1.08 & 0.95 & 0.95 \\
  &$\tauinter$ & -0.01 & 0.37 & 0.37 & 0.37 & 0.95 & 0.95 & 0.00 & 0.37 & 0.37 & 0.37 & 0.95 & 0.95 & 0.00 & 0.37 & 0.37 & 0.37 & 0.95 & 0.95 \\
  &$\tauLasso$ & -0.01 & 1.12 & 1.12 & 1.12 & 0.95 & 0.95 & -0.03 & 1.13 & 1.12 & 1.12 & 0.95 & 0.95 & -0.01 & 1.11 & 1.12 & 1.12 & 0.95 & 0.95 \\
  &$\tauLassos$ & 0.00 & 0.40 & 0.40 & 0.40 & 0.94 & 0.95 & 0.00 & 0.39 & 0.39 & 0.40 & 0.95 & 0.95 & 0.00 & 0.40 & 0.39 & 0.40 & 0.95 & 0.95 \\
  2&$\taustr$ & -0.02 & 2.14 & 2.11 & - & 0.94 & - & -0.03 & 2.13 & 2.11 & - & 0.95 & - & -0.04 & 2.12 & 2.11 & - & 0.95 & - \\
  &$\tauols$ & 0.04 & 1.63 & 1.64 & 1.64 & 0.95 & 0.95 & 0.04 & 1.66 & 1.64 & 1.65 & 0.94 & 0.94 & 0.04 & 1.64 & 1.64 & 1.65 & 0.95 & 0.95 \\
  &$\tauinter$ & 0.16 & 1.59 & 1.61 & 1.64 & 0.95 & 0.95 & 0.17 & 1.63 & 1.60 & 1.64 & 0.94 & 0.95 & 0.16 & 1.62 & 1.60 & 1.64 & 0.95 & 0.95 \\
  &$\tauLasso$ & 0.01 & 1.67 & 1.66 & 1.70 & 0.95 & 0.95 & 0.01 & 1.70 & 1.67 & 1.71 & 0.95 & 0.95 & 0.01 & 1.68 & 1.67 & 1.71 & 0.95 & 0.95 \\
  &$\tauLassos$ & 0.00 & 1.82 & 1.75 & 1.92 & 0.94 & 0.96 & 0.01 & 1.82 & 1.75 & 1.92 & 0.94 & 0.96 & 0.02 & 1.80 & 1.75 & 1.92 & 0.94 & 0.97 \\
  3&$\taustr$ & -0.01 & 1.21 & 1.17 & - & 0.94 & - & 0.00 & 1.18 & 1.17 & - & 0.95 & - & 0.04 & 1.19 & 1.17 & - & 0.95 & - \\
  &$\tauols$ & 0.00 & 0.20 & 0.20 & 0.20 & 0.94 & 0.95 & 0.00 & 0.20 & 0.20 & 0.20 & 0.94 & 0.94 & 0.00 & 0.20 & 0.20 & 0.20 & 0.94 & 0.94 \\
  &$\tauinter$ & -0.01 & 0.52 & 0.25 & 0.33 & 0.91 & 0.96 & 0.00 & 0.57 & 0.24 & 0.32 & 0.91 & 0.96 & 0.02 & 1.87 & 0.26 & 0.35 & 0.90 & 0.96 \\
  &$\tauLasso$ & 0.00 & 0.22 & 0.21 & 0.22 & 0.94 & 0.94 & 0.00 & 0.22 & 0.21 & 0.21 & 0.94 & 0.94 & 0.00 & 0.22 & 0.21 & 0.21 & 0.94 & 0.94 \\
  &$\tauLassos$ & 0.00 & 0.64 & 0.55 & 0.63 & 0.91 & 0.95 & 0.00 & 0.59 & 0.53 & 0.61 & 0.92 & 0.95 & 0.01 & 0.61 & 0.54 & 0.62 & 0.92 & 0.95 \\
   \hline
		\end{tabular}}
\begin{tablenotes}
\item Note: SD, standard deviation; SE, standard error; CP, coverage probability; \\
           \hspace{2cm} \hphantom{Note:} unadj, unadjusted variance estimator; adj, adjusted variance estimator; \\
           \hspace{2cm} \hphantom{Note:} -, not available.
\end{tablenotes}
\end{threeparttable}
\end{table}

Tables~\ref{equalsim} and \ref{equalsim2} present the simulation results for sample size $n$ being $200$ and $500$, respectively. Overall, the biases of the treatment effect estimators are negligible. The bias of $\tauinter$ tends to be large under Model $2$ when the sample size $n = 200$, and as the sample size increases, the bias tends to decrease.

For the first two models, the five treatment effect estimators behave similarly under different randomization methods. First, the four regression-adjusted estimators have smaller standard deviations than $\taustr$, which is consistent with the asymptotic results. Second, $\tauols$ and $\tauLasso$ are comparable, but $\tauols$ generally has slightly smaller standard deviations, as it only uses covariates truly related to the outcomes. The relation between $\tauinter$ and $\tauLassos$ is similar to that between $\tauols$ and $\tauLasso$. Third, under Model $1$, which has stratum-specific coefficients, $\tauinter$ and $\tauLassos$ outperform $\tauols$ and $\tauLasso$, as expected. Fourth, the unadjusted variance estimators perform well with a large sample size and a few strata, whereas under a small sample size, the unadjusted variance estimators tend to under-estimate the empirical variances. After adjusting for the degrees of freedom, the variance estimators can produce confidence intervals with coverage probabilities of approximately 95\%.

For model $3$, where the number of subjects in each stratum is small and the asymptotic theory might  have yet to step in, $\tauinter$ appears to have larger standard deviations than the other three regression-adjusted estimators ($\tauols$, $\tauLasso$, and $\tauLassos$), and sometimes, may even have larger standard deviations than $\taustr$. Meanwhile, $\tauLassos$, as a stratum-specific regression-adjusted estimator, still has efficiency gain compared with $\taustr$, thus exhibiting robustness. Moreover, the stratum-common regression-adjusted estimators $\tauols$ and $\tauLasso$ exhibit superior performance in this case. As for the variance estimators, the unadjusted variance estimators often, and sometimes severely, under-estimate the empirical variances, and this drawback can be addressed using the adjusted variance estimators.

\section{Clinical trial example}

The Nefazodone cognitive behavioral analysis system of psychotherapy (CBASP) trial was conducted to compare the efficacies of three treatments for chronic depression \citep{Keller2000}. In this section, we focus on two of the treatments, Nefazodone and the combination of Nefazodone and the cognitive behavioral-analysis system of psychotherapy (CBASP). The total number of patients was $440$, and the outcome of interest was the final score of the 24-item Hamilton rating scale for depression. We used the Nefazodone CBASP trial data solely for the purpose of generating synthetic data to illustrate the capability of different regression-adjusted estimators to improve efficiency. A detailed data generation process is given in the Appendix. We consider five regression-adjusted estimators with adjusted variance estimators, and the results are shown in Table~\ref{tab::syndata}.

\begin{table}[H]
\label{tab::syndata}
\centering
\caption{Estimates, 95\% confidence intervals, and variance reductions under simple randomization and stratified block randomization for synthetic Nefazodone CBASP trial data.\label{tab::syndata}}
\resizebox{\textwidth}{28mm}{
\begin{tabular}{l|cccc|cccc}
		\cline{1-9}
Randomization		& \multicolumn{4}{c}{Equal Allocation ($\pi = 1/2$)} & \multicolumn{4}{c}{Unequal Allocation ($\pi = 2/3$)} \\ \cline{2-9}
Methods &  Estimator & Estimate & 95\% CI & Variance Reduction&  Estimator & Estimate & 95\% CI & Variance Reduction \\ \cline{1-9}
                & $\taustr$ & -4.64 & (-5.41, -3.86) & ---                & $\taustr$ & -5.54 & (-6.43, -4.64) & --- \\
Simple        & $\tauols$   & -4.83 & (-5.54, -4.12) & 14.97\%     & $\tauols$ & -5.54 & (-6.39, -4.70) & 11.73\% \\
Randomization   & $\tauinter$ & -4.82 & (-5.54, -4.11) & 14.14\%    &  $\tauinter$ & -5.58 & (-6.42, -4.73) & 11.32\% \\
                & $\tauLasso$ & -4.77 & (-5.48, -4.07) & 16.88\%   &  $\tauLasso$ & -5.45 & (-6.27, -4.64) & 17.81\% \\
                & $\tauLassos$ & -4.92 & (-5.61, -4.23) & 19.47\%  &  $\tauLassos$ & -5.45 & (-6.23, -4.67) & 24.84\% \\
 \hline
                & $\taustr$ & -5.01 & (-5.84, -4.18) & ---               & $\taustr$ & -5.18& (-6.07, -4.29) & --- \\
Stratified      & $\tauols$ & -4.92 & (-5.71, -4.13) & 10.24\%     & $\tauols$ & -5.40 & (-6.22, -4.58) & 16.11\% \\
Block           & $\tauinter$ & -4.90 & (-5.68, -4.12) & 11.48\%   & $\tauinter$  & -5.42 & (-6.24, -4.59) & 14.95\%\\
Randomization   & $\tauLasso$ & -5.23 & (-6.01, -4.45) & 11.61\%  &  $\tauLasso$ & -5.22 & (-5.99, -4.46) & 25.91\% \\
                & $\tauLassos$ & -5.15 & (-5.88, -4.42) & 23.26\% &  $\tauLassos$ & -5.47 & (-6.22, -4.73) & 30.46\% \\
 \hline
\end{tabular}}
\begin{tablenotes}
\item Note: CI, confidence interval.
\end{tablenotes}
\end{table}

As shown in Table~\ref{tab::syndata}, all five treatment effect estimators suggest a negative effect of the combination treatment compared to the Nefazodone treatment under different randomization methods and allocations. Compared to $\taustr$, all other regression-adjusted estimators improve the efficiency, as measured by the variance reduction ranging from $10.24\%$ to $30.46\%$ . The Lasso-adjusted estimators $\tauLasso$ and $\tauLassos$ tend to have larger variance reductions than the OLS-adjusted estimators $\tauols$ and $\tauinter$, indicating the benefit in efficiency gain obtained using high-dimensional covariates. Moreover, $\tauLassos$ are even more efficient than $\tauLasso$, which is as expected because of the relatively large number of patients within each stratum.

\section{Discussion}

In this paper, we propose two Lasso-adjusted treatment effect estimators based on a general theory of regression adjustment for covariate-adaptive randomization. Both Lasso-adjusted treatment effect estimators are generally more efficient than the stratified difference-in-means estimator, and are robust against model misspecification and a small sample size.  Taking into account both asymptotic efficiency and finite sample performance, we recommend the stratum-common Lasso-adjusted estimator $\tauLasso$ for cases with  many small strata and the stratum-specific Lasso-adjusted estimator $\tauLassos$ for cases with a few large strata.

The Lasso-adjusted estimators assume a strict sparsity structure in the projection coefficients $\beta_{\proj}(a)$ and $\beta_{\proj}(b)$; that is, the numbers of nonzero elements of $\beta_{\proj}(a)$ and $\beta_{\proj}(b)$ are much smaller than the sample size. In practice, however, the projection coefficients  may exhibit different sparsity structures, such as group sparsity. In such cases,  Lasso can be replaced by other penalized (or regularized)  estimators, such as  the group Lasso~\citep{Yuan2006}, adaptive Lasso~\citep{Zou2006,HuangZhang2008}, elastic net~\citep{Zou2005}, SCAD~\citep{FanLi2001},  and MCP~\citep{Zhang2010mcp}, among many others. It would be interesting to outline the conditions under which the adjusted vectors obtained from these penalized regressions satisfy Assumption~\ref{assum::gen}, and to compare the efficiency of the resulting treatment effect estimators  with that of Lasso.

\bibliographystyle{apalike}
\bibliography{causal}


\appendix



\section{Useful Lemmas}

We first introduce the following lemmas that are useful for our proofs. We will give several additional lemmas during the proof of the main results.

\begin{lemma}
\label{lem::b3}
Under Assumptions~\ref{assum::Q}--\ref{assum::A}, let $V_i = f(Y_i(1),Y_i(0),B_i, \bx_i)$ for some measurable function   $f(\cdot)$  such that  $E( |V_i| ) < \infty$, then,
$$
\frac{1}{n} \sumi A_i V_i  \xrightarrow{P} \pi  E(V_1).
$$
\end{lemma}

\begin{lemma}
\label{lem::proportion}
Under Assumptions \ref{assum::Q}--\ref{assum::A}, we have
$$
\frac{\nt}{n} \xrightarrow{P} \pi, \quad \pik = \frac{\nkt}{\nk} \xrightarrow{P} \pi, \quad \frac{\nkt}{n}  \xrightarrow{P} \pi \pk, \quad \pnk = \frac{\nk}{n}  \xrightarrow{P} \pk,
$$
$$
\frac{\nc}{n} \xrightarrow{P} 1 - \pi, \quad \frac{\nkc}{\nk} \xrightarrow{P} 1 - \pi,  \quad \frac{\nkc}{n}  \xrightarrow{P} ( 1 -  \pi ) \pk.
$$
\end{lemma}

\citet{Bugni2019} obtained Lemma~\ref{lem::b3} for $V_i = f(Y_i(1),Y_i(0),B_i)$ (see Lemma C.4). Their proof can be easily generalized to $V_i = f(Y_i(1),Y_i(0),B_i, \bx_i)$. Lemma~\ref{lem::proportion} can be obtained directly from  the weak law of large numbers and the above Lemma~\ref{lem::b3}. We omit the proofs of these two lemmas.

\section{Proof of main results}

\subsection{Proof of Theorem~\ref{thm::gen}}

Before proving the theorem, we introduce the following lemma obtained from the proof of Lemma 7 in \citet{Ma2020Regression}.
\begin{lemma}
\label{lem::var}
Under Assumptions~\ref{assum::Q}--\ref{assum::A}, let $V_i = f(Y_i(1),Y_i(0),B_i, \bx_i)$ for some measurable function   $f(\cdot)$  such that  $E ( V_i^2 ) < \infty$, then,
$$
\sumk \pnk \cdot \frac{1}{\nkt} \sumik  A_i ( V_i - \bar{V}_{[k]1} )^2  \xrightarrow{P}  \sigma^2_{V_i - E(V_i \mid B_i ) },
$$
$$
\sumk \pnk \cdot \frac{1}{\nkc} \sumik  (1 - A_i )  ( V_i - \bar{V}_{[k]0} )^2  \xrightarrow{P}  \sigma^2_{V_i - E(V_i \mid B_i ) }.
$$
\end{lemma}

\begin{proof}[Proof of Theorem~\ref{thm::gen}]
Recall that
\begin{equation}
\label{eqn::taugen}
 \taugen = \sumk \pnk  \Big[  \big\{   \YkThat - ( \XkThat -  \XkT )^\T \betaTs   \big\}  - \big\{  \YkChat - ( \XkChat -  \XkC )^\T  \betaCs   \big\}  \Big] .
\end{equation}
It is easy to see that
$$
\XkT = \pik \XkThat + ( 1 - \pik ) \XkChat.
$$
Thus,
$$
\XkThat -  \XkT = ( 1 - \pik ) ( \XkThat - \XkChat), \quad \XkChat -  \XkC = - \pik  ( \XkThat - \XkChat).
$$
Taking them into \eqref{eqn::taugen}, we have
\begin{eqnarray}
\label{eqn::taugen1}
& & \taugen  \nonumber \\
& = & \sumk \pnk \Big[    \YkThat -  \YkChat  -  ( \XkThat -  \XkChat )^\T  \big \{ ( 1 - \pik ) \betaTs + \pik   \betaCs \big \} \Big]  \nonumber \\
& = & \sumk \pnk \Big[    \YkThat -  \YkChat  -  ( \XkThat -  \XkChat )^\T \beta_{[k]}^{*}  \Big]  -    \sumk \pnk (  \XkThat -  \XkChat )^\T \{ \hat \beta_{[k]}^{*} -  \beta_{[k]}^{*} \}, \nonumber \\
\end{eqnarray}
where
$$
\beta_{[k]}^{*} = (1 - \pi ) \beta_{[k]}(1) + \pi \beta_{[k]}(0), \quad \hat \beta_{[k]}^{*} = ( 1 - \pik ) \betaTs + \pik  \betaCs  .
$$
We will show that the first term in \eqref{eqn::taugen1} is asymptotically normal and the second term is asymptotically negligible.

The first term is the stratified difference-in-means estimator applied to the transformed outcomes $r_{i,\gen}(a)$, $a=0,1$, which satisfy
$$
E\{ r_{i,\gen}(1) - r_{i,\gen}(0) \} = \sumk \pk E\{ Y_i(1) - Y_i(0) \mid B_i = k \} = E \{ Y_i(1) - Y_i(0) \} =  \tau.
$$
Moreover, conditional on $B_i = k$,
\begin{eqnarray}
\label{eqn::second moment of ri}
 r_{i,\gen}(1) & = & Y_i(1) - \bx_i ^\T \beta_{[k]}^{*} \nonumber \\
 &=& (1 - \pi ) \{  Y_i(1) -   \bx_i ^\T  \beta_{[k]}(1)  \} + \pi \{  Y_i(0) -   \bx_i ^\T  \beta_{[k]}(0)  \} + \pi \{  Y_i(1) - Y_i(0) \}   \nonumber \\
 & = & ( 1 - \pi ) \r_{i, \gen}(1) + \pi  \r_{i, \gen}(0) +  \pi \{  Y_i(1) - Y_i(0) \} \nonumber \\
 & \in & \mathcal{L}_2,
\end{eqnarray}
where the last line is due to $\{ Y_i(1), Y_i(0) \} \in \mathcal{L}_2$ and $ \{  \r_{i, \gen}(1),  \r_{i, \gen}(0) \} \in \mathcal{L}_2 $. Similarly, $  r_{i,\gen}(0)  \in \mathcal{L}_2$.  Then, according to Proposition~\ref{prop::str}, the first term in \eqref{eqn::taugen1} is asymptotically normal with mean $\tau$ and variance $  \varsigma^2_{r_\gen}(\pi) + \varsigma^2_{H r_\gen} $. Thus,  for the asymptotic normality of $\taugen$ ,  it suffices to show that the second term in \eqref{eqn::taugen1} is asymptotically negligible.

For the second term, it holds that
\begin{eqnarray}
\label{eqn::second-term-gen}
&& (  \XkThat -  \XkChat )^\T \{ \hat \beta_{[k]}^{*} -  \beta_{[k]}^{*} \}  \nonumber \\
&= & (  \XkThat -  \XkChat )^\T \left\{ ( 1 - \pik ) \betaTs  - (1 - \pi ) \beta_{[k]}(1) \right\} +  (  \XkThat -  \XkChat )^\T \left\{  \pik  \betaCs  -  \pi  \beta_{[k]}(0) \right\} \nonumber \\
&  = &  (  \XkThat -  \XkChat )^\T \left[ ( 1 - \pik ) \left\{  \betaTs - \beta_{[k]}(1) \right\}  \right] +  (  \XkThat -  \XkChat )^\T \left\{ ( \pi - \pik ) \beta_{[k]}(1) \right\}  \nonumber \\
&& +  (  \XkThat -  \XkChat )^\T \left[  \pik \left\{  \betaCs  - \beta_{[k]}(0) \right\} \right] +  (  \XkThat -  \XkChat )^\T \left\{ (  \pik - \pi ) \beta_{[k]}(0) \right\}, \nonumber \\
& = &  ( \pi - \pik )   (  \XkThat -  \XkChat )^\T \left\{ \beta_{[k]}(1) - \beta_{[k]}(0) \right\} + o_P\left( \frac{ 1 }{ \surd{ n } }  \right),
\end{eqnarray}
where the last equality is because Assumption~\ref{assum::gen}.  The term $  (  \XkThat -  \XkChat )^\T \{ \beta_{[k]}(1) - \beta_{[k]}(0) \} $ is the difference-in-means estimator applied to the transformed covariates within stratum $k$,
$$ \{  \bx_i -  E ( \bx_i \mid B_i = k ) \}^\T  \{ \beta_{[k]}(1) - \beta_{[k]}(0) \}. $$
Recall the definition of the transformed outcomes $\r_{i,\gen}(a)$: conditional on $B_i = k$,
$$
\r_{i,\gen}(a) = Y_i(a)  -  \bx_i ^\T \beta_{[k]}(a).
$$
As we assume that both the potential outcomes $Y_i(a)  $ and the transformed outcomes $\r_{i,\gen}(a) $ belong to $ \mathcal{L}_2 $, then the transformed covariates within stratum $k$ satisfy
$$ \{  \bx_i -  E ( \bx_i \mid B_i = k ) \}^\T  \{ \beta_{[k]}(1) - \beta_{[k]}(0) \}  \in  \mathcal{L}_2. $$
Applying Proposition~\ref{prop::str} to the above transformed covariates within stratum $k$ (the maximum of its stratum-specific variance may be equal to zero, but it does not affect its asymptotic normality), we have
\begin{equation}
\label{eqn::second-term-gen2}
\surd{n}   (  \XkThat -  \XkChat )^\T \left\{ \beta_{[k]}(1) - \beta_{[k]}(0) \right\} = O_P (1).
\end{equation}
Taking \eqref{eqn::second-term-gen2} into \eqref{eqn::second-term-gen}, together with $   \pi - \pik = o_P(1) $ (Lemma~\ref{lem::proportion}), we have,
$$
 ( \pi - \pik )   (  \XkThat -  \XkChat )^\T \left\{ \beta_{[k]}(1) - \beta_{[k]}(0) \right\}  =  o_P\left( \frac{ 1 }{ \surd{ n } }  \right).
$$
Thus,
\begin{equation}
\label{eqn::xt-xc beta}
\surd{n}  (  \XkThat -  \XkChat )^\T \{ \hat \beta_{[k]}^{*} -  \beta_{[k]}^{*} \}  = o_P ( 1 ) .
\end{equation}
Therefore, the second term in \eqref{eqn::taugen1} is asymptotically negligible.


Next, we prove the consistency of the variance estimator.  By definition and simple calculation, we have, for $i \in [k]$,
\begin{eqnarray}
&& \hat r_{i,\gen}(1) - \frac{1}{\nkt} \sumik A_i \hat r_{i,\gen}(1)  \nonumber \\
& = & Y_i(1) - \bar{Y}_{[k]1} -  \big (  \bx_i - \bar{\bx}_{[k]1} \big ) ^\T \hat \beta_{[k]}^{*} \nonumber \\
&  =  & r_{i,\gen}(1) - \bar{r}_{[k]1,\gen} -  \big (  \bx_i - \bar{\bx}_{[k]1} \big ) ^\T \left\{ \hat \beta_{[k]}^{*} - \beta_{[k]}^{*} \right\} \nonumber \\
& = & r_{i,\gen}(1) - \bar{r}_{[k]1,\gen} - ( 1 - \pik )  \big (  \bx_i - \bar{\bx}_{[k]1} \big ) ^\T \left\{ \betaTs - \beta_{[k]}(1)  \right\} -  \nonumber \\
&&  \pik \big (  \bx_i - \bar{\bx}_{[k]1} \big ) ^\T \left\{ \betaCs - \beta_{[k]}(0)  \right\}  - ( \pi - \pik ) \big (  \bx_i - \bar{\bx}_{[k]1} \big ) ^\T \left\{ \beta_{[k]}(1) - \beta_{[k]}(0)  \right\}. \nonumber
\end{eqnarray}
In the following, we will deal with the sample variance of the above terms  separately. By Lemma~\ref{lem::var}, we have
\begin{equation}
\label{eqn::vargen1}
\sumk \pnk \cdot \frac{1}{\nkt} \sumik  A_i \left\{  r_{i,\gen}(1) - \bar{r}_{[k]1,\gen}  \right\}^2   \xrightarrow{P}  \sigma^2_{r_{i,\gen}(1) - E\{ r_{i,\gen}(1) \mid B_i \} }.
\end{equation}
As the covariates are uniformly bounded by $M$ (Assumption~\ref{assum::Q}), then by H\"older inequality, we have
\begin{eqnarray}
\label{eqn::vargen1later}
 | ( 1 - \pik )  \big (  \bx_i - \bar{\bx}_{[k]1} \big ) ^\T \left\{ \betaTs - \beta_{[k]}(1)  \right\} |  & \leq &  ||  \big (  \bx_i - \bar{\bx}_{[k]1} \big )  ||_{\infty} || \betaTs - \beta_{[k]}(1)   ||_1 \nonumber \\
& \leq & 2M || \betaTs - \beta_{[k]}(1)   ||_1 \xrightarrow{P} 0 ,
\end{eqnarray}
where the convergence in probability is due to Assumption~\ref{assum::gen}. Therefore,
\begin{eqnarray}
\label{eqn::vargen2}
&& \sumk \pnk \cdot \frac{1}{\nkt} \sumik  A_i  \left[ ( 1 - \pik )  \big (  \bx_i - \bar{\bx}_{[k]1} \big ) ^\T \left\{ \betaTs - \beta_{[k]}(1)  \right\} \right]^2  \nonumber \\
& \leq & 4 M^2 || \betaTs - \beta_{[k]}(1)   ||_1^2 \xrightarrow{P} 0.
\end{eqnarray}
Similarly,
\begin{eqnarray}
\label{eqn::vargen3}
&& \sumk \pnk \cdot \frac{1}{\nkt} \sumik  A_i  \left[ \pik \big (  \bx_i - \bar{\bx}_{[k]1} \big ) ^\T \left\{ \betaCs - \beta_{[k]}(0)  \right\}  \right]^2  \xrightarrow{P} 0.
\end{eqnarray}
For the last term, we have
\begin{eqnarray}
\label{eqn::vargen4}
&& \sumk \pnk \cdot \frac{1}{\nkt} \sumik  A_i \left[  ( \pi - \pik ) \big (  \bx_i - \bar{\bx}_{[k]1} \big ) ^\T \left\{ \beta_{[k]}(1) - \beta_{[k]}(0)  \right\}  \right]^2 \nonumber \\
 & \leq & \max_{k=1,\dots,K} ( \pi - \pik )^2 \cdot \sumk \pnk \cdot \frac{1}{\nkt} \sumik  A_i  \left[  \big (  \bx_i - \bar{\bx}_{[k]1} \big ) ^\T \left\{ \beta_{[k]}(1) - \beta_{[k]}(0)  \right\}  \right]^2 \nonumber \\
 & \xrightarrow{P} & 0,
\end{eqnarray}
where the convergence in probability is because of $ \pik   \xrightarrow{P}  \pi$ and  Lemma~\ref{lem::var} applied to $V_i = \{ \bx_i - E( \bx_i \mid B_i ) \}^\T \{ \beta_{[k]}(1) - \beta_{[k]}(0)  \} $ (Note that, $V_i \in \mathcal{L}_2$ because  both $Y_i(a) \in \mathcal{L}_2$ and $ \r_{i,\gen}(a) \in \mathcal{L}_2 $).

Combining \eqref{eqn::vargen1}--\eqref{eqn::vargen4} and using Cauchy--Schwarz inequality for the product terms, we have
\begin{eqnarray}
&&\sumk \pnk \cdot \frac{1}{\nkt} \sumik A_i \Big\{ \hat r_{i,\gen}(1)   - \frac{1}{\nkt} \sum_{j \in [k]} A_j \hat r_{j,\gen}(1)  \Big\}^2     \xrightarrow{P}  \sigma^2_{r_{i,\gen}(1) - E\{ r_{i,\gen}(1) \mid B_i \} }. \nonumber
\end{eqnarray}
Similarly,
\begin{eqnarray}
\sumk \pnk \cdot \frac{1}{\nkc} \sumik ( 1 - A_i )  \Big\{ \hat r_{i,\gen}(0)   - \frac{1}{\nkc} \sum_{j \in [k]} ( 1 - A_j ) \hat r_{j,\gen}(0)  \Big\}^2      \xrightarrow{P}  \sigma^2_{r_{i,\gen}(0) - E\{ r_{i,\gen}(0) \mid B_i \} }. \nonumber
\end{eqnarray}
Therefore,
\begin{equation}
\label{eqn::conver-vargen1}
\hat \varsigma^2_{r_\gen }(\pi)  \xrightarrow{P} \varsigma^2_{ r_\gen}(\pi) .
\end{equation}

To prove the consistency of $ \hat \varsigma^2_{H r_{\gen} } $, recall that
\begin{eqnarray}
\hat  \varsigma^2_{H r_{\gen}} & = & \sumk \pnk \bigg[ \Big\{  \frac{1}{\nkt} \sum_{j \in [k]} A_j \hat r_{j,\gen}(1) -   \frac{1}{\nt} \sumi  A_i  \hat r_{i,\gen}(1)  \Big\}  \nonumber \\
&& \quad \quad \quad \quad -  \Big\{ \frac{1}{\nkc} \sum_{j \in [k]} ( 1 - A_j ) \hat r_{j,\gen}(0) -  \frac{1}{\nc} \sumi  (1 - A_i )  \hat r_{i,\gen}(0)  \Big\} \bigg]^2. \nonumber
\end{eqnarray}
It suffices to show that
\begin{equation}
\label{eqn::hr1}
 \frac{1}{\nkt} \sum_{j \in [k]} A_j \hat r_{j,\gen}(1) -  \frac{1}{\nkc} \sum_{j \in [k]} ( 1 - A_j ) \hat r_{j,\gen}(0)  \xrightarrow{P} E\{  r_{i,\gen}(1) \mid B_i = k \} - E\{  r_{i,\gen}(0) \mid B_i = k \},
\end{equation}
\begin{equation}
\label{eqn::hr2}
\frac{1}{\nt} \sumi  A_i  \hat r_{i,\gen}(1)  -  \frac{1}{\nc} \sumi  (1 - A_i )  \hat r_{i,\gen}(0)  \xrightarrow{P} E\{  r_{i,\gen}(1)  \} - E\{  r_{i,\gen}(0)  \}.
\end{equation}
By definition and simple calculation, we have
\begin{eqnarray}
 \frac{1}{\nkt} \sum_{j \in [k]} A_j \hat r_{j,\gen}(1) & = & \frac{1}{\nkt} \sum_{j \in [k]} A_j  \big\{  Y_j(1) -   \bx_j   ^\T \hat \beta_{[k]}^{*} \big\} \nonumber \\
 & = &  \frac{1}{\nkt} \sum_{j \in [k]} A_j  r_{j, \gen}(1) -    \XkThat   ^\T  \big\{ \hat \beta_{[k]}^{*}  -  \beta_{[k]}^{*} \big\} \nonumber \\
 & = & \bar{r}_{[k]1, \gen}  -    \XkThat   ^\T  \big\{ \hat \beta_{[k]}^{*}  -  \beta_{[k]}^{*} \big\}. \nonumber
\end{eqnarray}
Similarly,
\begin{eqnarray}
 \frac{1}{\nkc} \sum_{j \in [k]} ( 1 - A_j ) \hat r_{j,\gen}(0) =  \bar{r}_{[k]0,\gen} -    \XkChat   ^\T  \big\{ \hat \beta_{[k]}^{*}  -  \beta_{[k]}^{*} \big\}. \nonumber
\end{eqnarray}
Therefore,
$$
 \frac{1}{\nkt} \sum_{j \in [k]} A_j \hat r_{j,\gen}(1) -  \frac{1}{\nkc} \sum_{j \in [k]} ( 1 - A_j ) \hat r_{j,\gen}(0)   = \bar{r}_{[k]1, \gen} - \bar{r}_{[k]0,\gen} -  (  \XkThat - \XkChat )   ^\T  \big\{ \hat \beta_{[k]}^{*}  -  \beta_{[k]}^{*} \big\}.
$$
We have shown in \eqref{eqn::xt-xc beta} that
$$ \surd{n}  (  \XkThat -  \XkChat )^\T \{ \hat \beta_{[k]}^{*} -  \beta_{[k]}^{*} \}  = o_P ( 1 ). $$
Applying Lemma~\ref{lem::b3} to $r_{i,\gen}(1)$, we have
$$  \bar{r}_{[k]1, \gen} - \bar{r}_{[k]0,\gen}    \xrightarrow{P} E\{  r_{i,\gen}(1) \mid B_i = k \} - E\{  r_{i,\gen}(0) \mid B_i = k \}.  $$
Therefore, statement \eqref{eqn::hr1} holds. To prove statement \eqref{eqn::hr2}, by definition and simple calculation, we have
\begin{eqnarray}
 \frac{1}{\nt} \sumi  A_i  \hat r_{i,\gen}(1)  & = &  \frac{1}{\nt} \sumk \sumik  A_i   \big\{  Y_i(1) -   \bx_i   ^\T \hat \beta_{[k]}^{*} \big\}  \nonumber \\
 & = &   \frac{1}{\nt} \sumk \sumik  A_i r_{i, \gen}(1) -  \sumk \frac{\nkt}{\nt}  \XkThat   ^\T  \big\{ \hat \beta_{[k]}^{*}  -  \beta_{[k]}^{*} \big\} \nonumber \\
 & = & \bar{r}_{1, \gen}  -   \sumk \frac{\nkt}{\nt}   \XkThat   ^\T  \big\{ \hat \beta_{[k]}^{*}  -  \beta_{[k]}^{*} \big\}. \nonumber
\end{eqnarray}
Similarly,
$$
\frac{1}{\nc} \sumi  (1 - A_i )  \hat r_{i,\gen}(0) = \bar{r}_{0, \gen}  -   \sumk \frac{\nkc}{\nc}   \XkChat   ^\T  \big\{ \hat \beta_{[k]}^{*}  -  \beta_{[k]}^{*} \big\}.
$$
Thus,
\begin{eqnarray}
&& \frac{1}{\nt} \sumi  A_i  \hat r_{i,\gen}(1)  -  \frac{1}{\nc} \sumi  (1 - A_i )  \hat r_{i,\gen}(0)  \nonumber \\
& = &  \bar{r}_{1, \gen}   -   \bar{r}_{0, \gen}  -  \sumk \frac{\nkc}{\nc}  ( \XkThat - \XkChat )   ^\T  \big\{ \hat \beta_{[k]}^{*}  -  \beta_{[k]}^{*} \big\} + \sumk \Big\{ \frac{\nkc}{\nc} -   \frac{\nkt}{\nt}  \Big\}  \XkThat   ^\T  \big\{ \hat \beta_{[k]}^{*}  -  \beta_{[k]}^{*} \big\} \nonumber \\
& = & E\{  r_{i,\gen}(1)  \} - E\{  r_{i,\gen}(0)  \} - \sumk \Big\{ \frac{\nkc}{\nc} -   \frac{\nkt}{\nt}  \Big\}  \XkThat   ^\T  \big\{ \hat \beta_{[k]}^{*}  -  \beta_{[k]}^{*} \big\}  + o_P(1), \nonumber
\end{eqnarray}
where the last equality is because of $  \bar{r}_{1, \gen}   -   \bar{r}_{0, \gen}   \xrightarrow{P}  E\{  r_{i,\gen}(1)  \} - E\{  r_{i,\gen}(0)  \} $, $\nkt / \nt  \xrightarrow{P} \pk$, and \eqref{eqn::xt-xc beta}. For statement \eqref{eqn::hr2}, it suffices to show that
$$
\sumk \Big\{ \frac{\nkc}{\nc} -   \frac{\nkt}{\nt}  \Big\}  \XkThat   ^\T  \big\{ \hat \beta_{[k]}^{*}  -  \beta_{[k]}^{*} \big\}  \xrightarrow{P}  0,
 $$
 which is implied by $ \nkc / \nc - \nkt / \nt  \xrightarrow{P}  0$ and
 \begin{eqnarray}
 &&  \XkThat   ^\T  \big\{ \hat \beta_{[k]}^{*}  -  \beta_{[k]}^{*} \big\}  \nonumber \\
 & = &  ( 1 - \pik )  \bar{\bx}_{[k]1} ^\T \left\{ \betaTs - \beta_{[k]}(1)  \right\} -   \pik  \bar{\bx}_{[k]1}  ^\T \left\{ \betaCs - \beta_{[k]}(0)  \right\}  - ( \pi - \pik )  \bar{\bx}_{[k]1}  ^\T \left\{ \beta_{[k]}(1) - \beta_{[k]}(0)  \right\}, \nonumber \\
 &  \xrightarrow{P} & 0. \nonumber
 \end{eqnarray}
The above convergence in probability is obtained by similar arguments as \eqref{eqn::vargen1later}--\eqref{eqn::vargen4}.

Finally, we study the minimizer of  the asymptotic variance of $\taugen$. Let $  \tilde \bx_i = \bx_i - E ( \bx_i \mid B_i )$, $ \tilde Y_i(a)  = Y_i(a) - E\{  Y_i(a) \mid B_i \}$, and let $\Sigma_{[k] \bx Y(a)} = E [  \tilde \bx_i  \tilde Y_i(a)   \mid B_i = k ]$ be the stratum-specific covariance of $\bx_i$ and $Y_i(a)$ in stratum $k$. Let $\tilde r_{i,\gen} (a) =  r_{i, \gen} (a) - E \{  r_{i,\gen}(a) \mid B_i \}$, then $E \{ \tilde r_{i,\gen} (a) \} = 0 $.
By definition and simple calculation, we have
\begin{eqnarray}
\sigma^2_{\tilde r_\gen (a)}  & = & E  \{ \tilde r_{i,\gen} (a) \}^2 \nonumber \\
& = & \sumk \pk  E \left( \left[  Y_{i}(a) - E\{ Y_i(a) \mid B_i = k \}  - \big\{ \bx_i - E( \bx_i \mid B_i = k ) \big\} ^\T \beta_{[k]}^{*}    \right ]^2 \mid B_i = k \right)  \nonumber \\
& = & \sigma^2_{\tilde Y(a)} + \sumk \pk ( \beta_{[k]}^{*} ) ^\T \Sigma_{[k] \bx \bx}  ( \beta_{[k]}^{*} ) - 2 \sumk \pk ( \beta_{[k]}^{*} ) ^\T \Sigma_{[k] \bx  Y(a)}. \nonumber
\end{eqnarray}
Therefore,
\begin{eqnarray}
\label{eqn::intervargen1}
 \varsigma^2_{r_\gen }(\pi) & = & \frac{1}{\pi} \sigma^2_{\tilde r_{\gen}(1) } + \frac{1}{1 - \pi} \sigma^2_{ \tilde r_{\gen}(0) }  \nonumber \\
  & = & \varsigma^2_{Y}(\pi) +  \frac{1}{\pi } \sumk \pk  ( \beta_{[k]}^{*} ) ^\T \Sigma_{[k] \bx  \bx} ( \beta_{[k]}^{*} ) + \frac{1}{1 - \pi } \sumk \pk  ( \beta_{[k]}^{*} ) ^\T \Sigma_{[k] \bx \bx} ( \beta_{[k]}^{*} )  \nonumber \\
 && - \frac{2}{\pi } \sumk \pk ( \beta_{[k]}^{*} ) ^\T   \Sigma_{[k] \bx Y(1)}  - \frac{2}{ 1 - \pi } \sumk \pk ( \beta_{[k]}^{*} ) ^\T   \Sigma_{[k] \bx Y(0)}  \nonumber \\
 & = &  \varsigma^2_{Y}(\pi)  +   \frac{1}{\pi ( 1 - \pi ) } \sumk \pk ( \beta_{[k]}^{*} ) ^\T \Sigma_{[k] \bx \bx} ( \beta_{[k]}^{*} ) \nonumber \\
 && - \frac{2}{ \pi ( 1 - \pi ) } \sumk \pk ( \beta_{[k]}^{*} ) ^\T \{ ( 1 - \pi )  \Sigma_{[k] \bx Y(1)} + \pi  \Sigma_{[k] \bx Y(0)} \}.
\end{eqnarray}
Since for $a=0,1$,
$$
E\{ r_{i, \gen} (a) \mid B_i = k \}  = E\{ Y_i(a) \mid B_i = k \}  - \left\{ E( \bx_i \mid B_i = k )  \right\} ^\T \beta_{[k]}^{*},
$$
\begin{eqnarray}
E\{ r_{i,\gen}(a) \} & = &  \sumk \pk \left[ E\{ Y_i(a) \mid B_i = k \}  - \left\{ E( \bx_i \mid B_i = k )  \right\} ^\T \beta_{[k]}^{*} \right] \nonumber \\
& = & E\{ Y_i(a) \} - \sumk \pk \left\{ E( \bx_i \mid B_i = k )  \right\} ^\T \beta_{[k]}^{*}, \nonumber
\end{eqnarray}
then,
\begin{eqnarray}
&& \left[ E\{ r_{i,\gen}(1) \mid B_i = k \} - E\{ r_{i,\gen}(1) \} \right] - \left[ E\{ r_{i,\gen}(0) \mid B_i = k \} - E\{ r_{i,\gen}(0) \} \right]  \nonumber \\
& = &  \left[ E\{ Y_i(1) \mid B_i = k \} - E\{ Y_i(1) \} \right] - \left[ E\{ Y_i(0) \mid B_i = k \} - E\{ Y_i(0) \} \right]. \nonumber
\end{eqnarray}
Therefore,
\begin{equation}
\label{eqn::intervargen2}
 \varsigma^2_{H r_\gen} =  \varsigma^2_{HY}.
\end{equation}
Combing \eqref{eqn::intervargen1} and \eqref{eqn::intervargen2}, the asymptotic variance of $\taugen$ satisfies
\begin{eqnarray}
\label{eqn::genvar-form}
&&  \varsigma^2_{ r_\gen}(\pi) + \varsigma^2_{H r_\gen}  \nonumber \\
 & = &   \varsigma^2_{Y}(\pi)  +   \varsigma^2_{HY}  +  \frac{1}{\pi ( 1 - \pi ) } \sumk \pk ( \beta_{[k]}^{*} ) ^\T \Sigma_{[k] \bx \bx} ( \beta_{[k]}^{*} ) \nonumber \\
 && - \frac{2}{ \pi ( 1 - \pi ) } \sumk \pk ( \beta_{[k]}^{*} ) ^\T \{ ( 1 - \pi )  \Sigma_{[k] \bx Y(1)} + \pi  \Sigma_{[k] \bx Y(0)} \}.
\end{eqnarray}
Now, we can obtain its minimizer.
\begin{itemize}
\item[(1)] Under the constraint that $\beta_{[k]}(a) = \beta(a)$ for $k= 1,\dots,K$, $a = 0, 1$. In this case, $ \beta_{[k]}^{*}  = ( 1 - \pi ) \beta(1) + \pi \beta(0)$, denoted by $\beta^* $, and equation \eqref{eqn::genvar-form} is reduced to
\begin{eqnarray}
&&  \varsigma^2_{ r_\gen}(\pi) + \varsigma^2_{H r_\gen}  \nonumber \\
 & = &   \varsigma^2_{Y}(\pi)  +   \varsigma^2_{HY}  +  \frac{1}{\pi ( 1 - \pi ) }  ( \beta^*)^\T   \Sigma_{\tilde \bx  \tilde \bx} ( \beta^* )  - \frac{2}{ \pi ( 1 - \pi ) }  (\beta^*)^\T \{ ( 1 - \pi )  \Sigma_{\tilde \bx \tilde Y(1)} + \pi  \Sigma_{\tilde \bx \tilde Y(0)} \}. \nonumber
\end{eqnarray}
Taking derivative with respect to $\beta^*$ and setting it to zero, we can obtain the minimizer
$$
(1 - \pi ) \Sigma_{\tilde \bx  \tilde \bx}^{-1}  \Sigma_{\tilde \bx \tilde Y(1)} + \pi  \Sigma_{\tilde \bx  \tilde \bx}^{-1}  \Sigma_{\tilde \bx \tilde Y(0)} = ( 1 - \pi ) \beta_{\proj}(1) + \pi \beta_{\proj} (0).
$$
Clearly, $\beta_{[k]}(a) = \beta_{\proj}(a)$ corresponds to this minimizer.

\item[(1)] Without constraint. Taking derivatives with respect to $ \beta_{[k]}^{*}$ and setting them to zero, we can obtain the minimizer
$$
  (1 - \pi ) \Sigma_{[k] \bx  \bx}^{-1}  \Sigma_{[k] \bx  Y(1)} + \pi  \Sigma_{[k] \bx   \bx}^{-1}  \Sigma_{[k] \bx  Y(0)} = ( 1 - \pi ) \beta_{[k] \proj}(1) + \pi \beta_{[k] \proj} (0).
$$
Clearly, $\beta_{[k]}(a) = \beta_{[k]\proj}(a)$ corresponds to this minimizer.
\end{itemize}

\end{proof}

\subsection{Proof of Theorem~\ref{thm::betaLasso}}

Before proving the theorem, we introduce the following lemma which provides concentration inequalities for stratum-specific sample means under covariate-adaptive randomization.
\begin{lemma}
\label{lem::concentration}
Under Assumptions~\ref{assum::Q}, \ref{assum::A}, and \ref{assum::sub-Gaussian}, we have the following concentration inequalities:
$$
|| \XkThat - E ( \bx_i \mid B_i = k ) ||_{\infty} = O_P\left\{   \bigg( \frac{\log p}{n} \bigg)^{1/2} \right\}, \quad ||  \XkThat - \XkT  ||_{\infty} = O_P\left\{   \bigg( \frac{\log p}{n} \bigg)^{1/2} \right\},
$$
$$
|| \XkChat - E ( \bx_i \mid B_i = k ) ||_{\infty} = O_P\left\{   \bigg( \frac{\log p}{n} \bigg)^{1/2} \right\}, \quad ||  \XkChat - \XkC  ||_{\infty} = O_P\left\{   \bigg( \frac{\log p}{n} \bigg)^{1/2} \right\}.
$$
\end{lemma}
The proof of Lemma~\ref{lem::concentration} will be given in Section~\ref{sec::proof-lemma}. Now, we can prove Theorem~\ref{thm::betaLasso}.

\begin{proof}[Proof of Theorem~\ref{thm::betaLasso}]
We will only prove that
$$
|| \betaLassoT - \beta_{\proj}(1) ||_1 = O_P\left\{   s \bigg( \frac{ M_n \log p}{n} \bigg)^{1/2} \right\},
$$
as the proof for the counterpart of the control  is similar. Recall that,
\begin{eqnarray*}
 \betaLassoT & = &    \argmin_{ \beta } \frac{1}{ \nt} \sumk \sumik A_i \left\{ Y_i - \YkThat -  (\bx_i - \XkThat )^\T \beta \right\}^2 + \lambda_1 || \beta ||_1 \\
 & = &  \argmin_{ \beta } \frac{1}{ \nt} \sumk \sumik A_i \left\{ Y_i(1) - \YkThat -  (\bx_i - \XkThat )^\T \beta \right\}^2 + \lambda_1 || \beta ||_1.
\end{eqnarray*}
By the definition of minimizer, we have
\begin{eqnarray}
\label{eqn::basic}
 && \frac{1}{ \nt} \sumk \sumik A_i \left\{ Y_i(1) - \YkThat -  (\bx_i - \XkThat )^\T \betaLassoT \right\}^2 + \lambda_1 || \betaLassoT ||_1 \nonumber \\
 & \leq &   \frac{1}{ \nt} \sumk \sumik A_i \left\{ Y_i(1) - \YkThat -  (\bx_i - \XkThat )^\T \beta_{\proj}(1) \right\}^2 + \lambda_1 || \beta_{\proj}(1) ||_1,
\end{eqnarray}
where $\beta_{\proj}(1)$ is the projection coefficient defined by
\begin{equation}
\label{eqn::tildebeta}
\beta_{\proj} (1) = \argmin_{ \beta } E \big [  Y_i(1) - E \{ Y_i(1) | B_i \} -  \{ \bx_i - E ( \bx_i | B_i )  \} ^\T \beta  \big ]^2,  \nonumber
\end{equation}
Recall that,
$$
 \r_i(1) = Y_i(1)   -  \bx_i ^\T \beta_{\proj}(1).
$$
Then,
$$
\r_i(1) - \bar{ \r }_{[k]1} = Y_i(1) - \YkThat -  (\bx_i - \XkThat )^\T  \beta_{\proj}(1) , \quad  i \in [k].
$$
Thus,
\begin{eqnarray}
\label{eqn::left}
&&  \frac{1}{ \nt} \sumk \sumik A_i \left\{ Y_i(1) - \YkThat -  (\bx_i - \XkThat )^\T \betaLassoT \right\}^2  \nonumber \\
& = &  \frac{1}{ \nt} \sumk \sumik A_i \left[  \r_i(1) - \bar{ \r }_{[k]1}  -  (\bx_i - \XkThat )^\T \left\{ \betaLassoT - \beta_{\proj}(1) \right\} \right]^2 \nonumber \\
& =&     \left\{  \betaLassoT -  \beta_{\proj}(1)  \right\}^\T \left\{  \frac{1}{ \nt} \sumk \sumik A_i  (\bx_i - \XkThat )  (\bx_i - \XkThat )^\T \right\}  \left\{  \betaLassoT -  \beta_{\proj}(1)  \right\}   \nonumber \\
&&  - \frac{2}{ \nt}  \sumk \sumik A_i  (\bx_i - \XkThat )^\T  \left \{ \r_i(1)  - \bar{\r}_{[k]1} \right \}  \left\{  \betaLassoT -  \beta_{\proj}(1)  \right\}   \nonumber \\
&& +  \frac{1}{ \nt}   \sumk \sumik A_i  \{ \r_i(1) - \bar{\r}_{[k]1} \}^2 ,
\end{eqnarray}
and
\begin{eqnarray}
\label{eqn::right}
 \frac{1}{ \nt} \sumk \sumik A_i \left\{ Y_i(1) - \YkThat -  (\bx_i - \XkThat )^\T \beta_{\proj}(1) \right\}^2  =  \frac{1}{ \nt}   \sumk \sumik A_i  \left\{ \r_i(1) - \bar{\r}_{[k]1} \right \}^2.
\end{eqnarray}
Taking \eqref{eqn::left} and \eqref{eqn::right} into \eqref{eqn::basic} and let $\bh =  \betaLassoT -  \beta_{\proj}(1)$, we have
\begin{eqnarray}
&&  \bh^\T \left\{  \frac{1}{ \nt} \sumk \sumik A_i  (\bx_i - \XkThat )  (\bx_i - \XkThat )^\T \right\}  \bh + \lambda_1 || \betaLassoT ||_1  \nonumber \\
&  \leq &  \frac{2 }{ \nt}  \sumk \sumik A_i  (\bx_i - \XkThat )^\T  \{ \r_i(1) - \bar{\r}_{[k]1} \}  \bh  + \lambda_1 || \beta_{\proj}(1) ||_1. \nonumber
\end{eqnarray}
Let
\begin{equation}
\label{eqn::hatSxx}
S_{\tilde \bx \tilde \bx  }(1) = \frac{1}{\nt} \sumk \sumik A_i ( \bx_i - \XkThat ) ( \bx_i - \XkThat )^\T,  \nonumber
\end{equation}
\begin{equation}
\label{eqn::hatSxeps}
S_{\tilde \bx \r(1) } = \frac{1}{\nt} \sumk \sumik A_i ( \bx_i - \XkThat ) \{  \r_i(1) - \bar{\r}_{[k]1} \}, \nonumber
\end{equation}
then
\begin{eqnarray}
\label{eqn::basic2}
 \bh^\T  S_{\tilde \bx \tilde \bx  }(1)  \bh + \lambda_1 || \betaLassoT ||_1  \leq  2 S_{\tilde \bx \r(1) } ^\T \bh + \lambda_1 || \beta_{\proj}(1) ||_1 \leq  || 2 S_{\tilde \bx \r(1)  } ||_{\infty}  || \bh ||_1 + \lambda_1 || \beta_{\proj}(1) ||_1,
\end{eqnarray}
where the last inequality is due to  H{\" o}eder inequality. Consider the following event
\begin{equation}
\mathcal{E}_1 = \{ || 2 S_{\tilde \bx \r(1)  } ||_{\infty}  \leq \lambda_1 / 2 \}
\end{equation}
\begin{lemma}
\label{lem::stochastic}
If $\{ r_{i}(1), r_{i}(0) \} \in \mathcal{L}_2 $ and Assumptions~\ref{assum::Q}--\ref{assum::A}, and \ref{assum::sub-Gaussian}--\ref{assum::sparsity} hold, then
$$P( \mathcal{E}_1 ) \geq 1 - \frac{K}{M_n}  \rightarrow 1.$$
\end{lemma}
The proof of Lemma~\ref{lem::stochastic} will be given in Section~\ref{sec::proof-lemma}. To proceed, conditional on $\mathcal{E}_1 $, we have
\begin{equation}
 \bh^\T  S_{\tilde \bx \tilde \bx  }(1)  \bh + \lambda_1 || \betaLassoT ||_1  \leq  \lambda_1 || \bh ||_1 /2   +  \lambda_1 || \beta_{\proj}(1) ||_1.
\end{equation}
Therefore,
\begin{equation}\label{eqn::bhsbh}
2 \bh^\T  S_{\tilde \bx \tilde \bx  }(1)  \bh + 2 \lambda_1 || \betaLassoT ||_1 \leq  \lambda_1 || \bh ||_1 +  2 \lambda_1 || \beta_{\proj}(1) ||_1.
\end{equation}
Recall that, for vector $u$, $u_{S} = (u_j, j \in S)^\T$. Using triangle inequality, we have
\begin{eqnarray}
|| \betaLassoT ||_1  & = &  || [\betaLassoT]_{S} ||_1 + || [\betaLassoT]_{S^c}  ||_1 \nonumber \\
& \geq &  ||  [ \beta_{\proj} (1) ]_{S} ||_1 -  || [\betaLassoT - \beta_{\proj} (1) ]_{S}  ||_1 + || [\betaLassoT]_{S^c}  ||_1 \nonumber \\
& = &  ||  [ \beta_{\proj} (1) ]_{S} ||_1 -  || \bh_{S}  ||_1 + || [\betaLassoT]_{{S}^c}  ||_1, \nonumber \\
& = &  ||   \beta_{\proj} (1) ||_1 -  || \bh_{S}  ||_1 + || [\betaLassoT]_{{S}^c}  ||_1, \nonumber
\end{eqnarray}
where $S  = \{ j \in \{ 1, \dots, p \}: \beta_{j,\proj}(1) \neq 0 \ \text{or} \ \beta_{j,\proj}(0) \neq 0 \}$. Thus,
\begin{equation}\label{eqn::betahat-beta}
|| \betaLassoT ||_1 -   ||   \beta_{\proj} (1) ||_1 \geq  -  || \bh_{S}  ||_1 + || [\betaLassoT]_{{S}^c}  ||_1 =  -  || \bh_{S}  ||_1 + || \bh_{S^c} ||_1,
\end{equation}
where the last equality is because of $ [ \beta_{\proj} (1) ]_{S^c} = 0$ and $ || [\betaLassoT]_{S^c}  ||_1 = || \bh_{S^c} ||_1$. Moreover,
\begin{equation}\label{eqn::bh1}
 || \bh ||_1 =  || \bh_{S} ||_1 +  || \bh_{S^c}  ||_1 .
\end{equation}
Taking \eqref{eqn::betahat-beta} and \eqref{eqn::bh1} into \eqref{eqn::bhsbh} yields
\begin{equation}
\label{eqn::basic3}
2  \bh^\T  S_{\tilde \bx \tilde \bx  }(1)  \bh + \lambda_1 || \bh_{S^c} ||_1 \leq 3 \lambda_1 || \bh_{S} ||_1.
\end{equation}
As $ \bh^\T  S_{\tilde \bx \tilde \bx  }(1)  \bh \geq 0$, we have
\begin{equation}
\label{eqn::cone}
 || \bh_{S^c} ||_1 \leq 3  || \bh_{S} ||_1.
\end{equation}


\begin{lemma}
\label{lem::RE}
Let $\mathcal{C} = \{  \bh \in R^p:  || \bh_{S^c} ||_1 \leq 3 || \bh_{S}  ||_1 \}$. Under Assumptions~\ref{assum::sub-Gaussian} and \ref{assum::sparsity}, there exits a constant $c_{\min}$ not depending on $n$, such that for the event $\mathcal{E}_2 = \{ \bh^\T  S_{\tilde \bx \tilde \bx  }(1)  \bh \geq c_{\min} || \bh_{S} ||_2^2 \} $ where $\bh \in \mathcal{C}$, we have
$$
P\left( \mathcal{E}_2  \right) \rightarrow 1.
$$
\end{lemma}
The proof of Lemma~\ref{lem::RE} will be given in Section~\ref{sec::proof-lemma}. To proceed, conditional on the event $\mathcal{E}_2$ and by \eqref{eqn::basic3}, we have
$$
 c_{\min} || \bh_{S} ||_2^2 \leq \bh^\T  S_{\tilde \bx \tilde \bx  }(1)  \bh \leq    \frac{3}{2}  \lambda_1 || \bh_{S} ||_1 \leq \frac{3}{2} \surd{s} \lambda_1 || \bh_{S} ||_2,
$$
where $s = |S| $ and the last inequality is due to  Cauchy--Schwarz inequality.  Therefore,
\begin{equation}\label{eqn::bhslambda}
|| \bh_{S} ||_2 \leq \frac{3}{2 c_{\min}}  \surd{s} \lambda_1.
\end{equation}
Combining \eqref{eqn::cone} and \eqref{eqn::bhslambda}, we have
$$
|| \bh ||_1 =  || \bh_{S^c} ||_1 +   || \bh_{S} ||_1 \leq 4   || \bh_{S} ||_1 \leq 4 \surd{s} || \bh_{S}  ||_2 \leq \frac{6}{c_{\min}}  s \lambda_1.
$$
Therefore, conditional on $\mathcal{E}_1$ and $\mathcal{E}_2$, we have
$$
|| \betaLassoT - \beta_{\proj}(1)  ||_1 =  || \bh ||_1  \leq \frac{6}{c_{\min}}  s \lambda_1.
$$
Combining Lemmas~\ref{lem::stochastic} and \ref{lem::RE}, and  Assumption~\ref{assum::lambda} yield
$$
|| \betaLassoT - \beta_{\proj}(1)  ||_1 = O_P \left\{   s \bigg( \frac{ M_n \log p}{n} \bigg)^{1/2} \right\}.
$$

\end{proof}

\subsection{Proof of Theorem~\ref{thm::Lasso}}

\begin{proof}
For the asymptotic normality of $\tauLasso$ and consistency of the variance estimator, it suffices to show that the stratum-common Lasso-adjusted vectors $\betaLassoT$ and $\betaLassoC$ satisfy Assumption~\ref{assum::gen} with $\hat \beta_{[k]}(1) = \betaLassoT$, $\hat \beta_{[k]}(0) = \betaLassoC$,  $\beta_{[k]}(1) = \beta_{\proj} (1)$, and $\beta_{[k]}(0) = \beta_{\proj}(0)$, $k=1,\dots,K$. According to Theorem~\ref{thm::betaLasso} and the sparsity Assumption~\ref{assum::sparsity}, we have
\begin{equation}
\label{eqn::consis-betaLasso}
|| \hat \beta_{\Lasso}(a) -  \beta_{\proj}(a)  ||_1 = O_P \left\{   s \bigg( \frac{ M_n \log p}{n} \bigg)^{1/2} \right\} = o_P(1), \quad a = 0,1.
\end{equation}
By Lemma~\ref{lem::concentration}
$$
 ||  \XkThat - \XkT  ||_{\infty} = O_P\left\{   \bigg( \frac{\log p}{n} \bigg)^{1/2} \right\},\quad  ||  \XkChat - \XkT  ||_{\infty} = O_P\left\{   \bigg( \frac{\log p}{n} \bigg)^{1/2} \right\}.
$$
Thus,
$$
 ||  \XkThat - \XkChat  ||_{\infty} = O_P\left\{   \bigg( \frac{\log p}{n} \bigg)^{1/2} \right\}.
$$
By H\"older inequality and \eqref{eqn::consis-betaLasso}, we have, for $a=0,1$  and $k=1,\dots,K$,
\begin{eqnarray}
&& | \surd{n} (  \XkThat -  \XkChat )^\T  \left\{  \hat \beta_{\Lasso}(a) - \beta_{\proj} (a) \right\}   | \nonumber \\
& \leq & \surd{n} || \XkThat -  \XkChat  ||_{\infty} \cdot  ||  \left\{  \hat \beta_{\Lasso}(a)  - \beta_{\proj}(a) \right\} ||_1 \nonumber \\
& = & O_P\left\{ \surd{n}  \bigg( \frac{\log p}{n} \bigg)^{1/2} \right\} \cdot O_P\left\{ s \bigg( \frac{M_n \log p}{n} \bigg)^{1/2} \right\} \nonumber \\
& = & O_P\left( \frac{\surd{M_n} s \log p}{ \surd{n} }  \right), \nonumber \\
& = & o_P(1), \nonumber
\end{eqnarray}
where the last equality is because of Assumption~\ref{assum::sparsity}. Therefore, $\betaLassoT$ and $\betaLassoC$ satisfy Assumption~\ref{assum::gen} with $\beta_{[k]}(1) = \beta_{\proj} (1)$ and $\beta_{[k]}(0) = \beta_{\proj}(0)$.

Next, we compare the asymptotic variance of $\tauLasso$ and $\taustr$. Denote $\tilde r_i(a) = r_i(a) - E \{ r_i(a) \mid B_i \}$, and $\tilde Y_i(a) = Y_i(a) - E \{ Y_i(a) \mid B_i \}$,  $a=0,1$. Simple calculation gives
\begin{eqnarray}
\sigma^2_{\tilde r(a)} & = & \Var[ Y_i(a) - E \{ Y_i(a) | B_i \} - ( \bx_i - E \{ \bx_i | B_i \} )^\T \beta_{\proj}^ {*} ]  \nonumber \\
& = & \sigma^2_{\tilde Y(a)} +  ( \beta_{\proj}^ {*})^\T \Sigma_{\tilde \bx \tilde \bx} (\beta_{\proj}^ {*}) - 2 ( \beta_{\proj}^ {*})^\T \Sigma_{\tilde \bx \tilde Y(a)} \nonumber \\
& = &  \sigma^2_{\tilde Y(a)} +  ( \beta_{\proj}^ {*})^\T \Sigma_{\tilde \bx \tilde \bx} (\beta_{\proj}^ {*}) - 2 ( \beta_{\proj}^ {*})^\T \Sigma_{\tilde \bx \tilde \bx} \beta_{\proj}(a), \nonumber
\end{eqnarray}
where $\tilde \bx = \bx - E( \bx \mid B )$, and the last equality is because of $\beta_{\proj}(a) = \Sigma_{\tilde \bx \tilde \bx}^{-1} \Sigma_{\tilde \bx \tilde Y(a)}$. Therefore,
\begin{eqnarray}
\label{eqn::intervar1}
&&  \varsigma^2_{ r}(\pi) - \varsigma^2_{ Y}(\pi) \nonumber \\
  & = & \frac{ \sigma^2_{\tilde r (1)}  - \sigma^2_{\tilde Y(1)}  }{\pi} +  \frac{\sigma^2_{\tilde r (0)}  - \sigma^2_{\tilde Y(0)}  }{ 1 - \pi } \nonumber \\
 & = & \frac{1}{\pi }  ( \beta_{\proj}^ {*})^\T \Sigma_{\tilde \bx \tilde \bx} (\beta_{\proj}^ {*}) + \frac{1}{1 - \pi } ( \beta_{\proj}^ {*})^\T \Sigma_{\tilde \bx \tilde \bx} (\beta_{\proj}^ {*})  - \frac{2}{\pi } ( \beta_{\proj}^ {*})^\T   \Sigma_{\tilde \bx \tilde \bx}  \beta_{\proj}(1) \nonumber \\
 && - \frac{2}{ 1 - \pi }  ( \beta_{\proj}^ {*})^\T   \Sigma_{\tilde \bx \tilde \bx}  \beta_{\proj}(0) \nonumber \\
 & = &   \frac{1}{\pi ( 1 - \pi ) }  ( \beta_{\proj}^ {*})^\T \Sigma_{\tilde \bx \tilde \bx} (\beta_{\proj}^ {*}) - \frac{2}{ \pi ( 1 - \pi ) } ( \beta_{\proj}^ {*})^\T \Sigma_{\tilde \bx \tilde \bx} \{ ( 1 - \pi ) \beta_{\proj}(1) + \pi \beta_{\proj}(0) \} \nonumber \\
 & = & -  \frac{1}{\pi ( 1 - \pi ) }  ( \beta_{\proj}^ {*})^\T \Sigma_{\tilde \bx \tilde \bx} ( \beta_{\proj}^ {*}),
\end{eqnarray}
where  the last equality is due to $\beta_{\proj}^ {*}  = ( 1 - \pi ) \beta_{\proj}(1) + \pi \beta_{\proj}(0) $. By similar arguments as the proof of Theorem~\ref{thm::gen}, we have
\begin{equation}
\label{eqn::intervar2}
 \varsigma^2_{Hr} =  \varsigma^2_{HY}.
\end{equation}
Combing \eqref{eqn::intervar1} and \eqref{eqn::intervar2}, the difference of the asymptotic variances of $\tauLasso$ and $\taustr$ is
\begin{eqnarray}
\Delta & = &   \{ \varsigma^2_{ r}(\pi) + \varsigma^2_{Hr}  \} - \{ \varsigma^2_{ Y}(\pi) + \varsigma^2_{HY}  \}  \nonumber \\
& = &  \varsigma^2_{ r}(\pi)  - \varsigma^2_{ Y}(\pi)  \nonumber \\
& = &  -  \frac{1}{\pi ( 1 - \pi ) }  ( \beta_{\proj}^ {*})^\T \Sigma_{\tilde \bx \tilde \bx} ( \beta_{\proj}^ {*} ) \leq 0. \nonumber
\end{eqnarray}

\end{proof}

\subsection{Proof of Corollary~\ref{cor::betaLasso}}
\begin{proof}
As Theorem~\ref{thm::betaLasso} holds for $K=1$, the conclusion follows immediately by applying Theorem~\ref{thm::betaLasso} (with $K = 1$) to each stratum $k$ separately.
\end{proof}

\subsection{Proof of Theorem~\ref{thm::Lasso2}}
\begin{proof}

Similar to the proof of Theorem~\ref{thm::Lasso}, for the  asymptotic normality of $\tauLassos$ and consistency of the variance estimator, it suffices to show that the stratum-specific Lasso-adjusted vectors $\betaLassoTs$ and $\betaLassoCs$ satisfy Assumption~\ref{assum::gen} with $\hat \beta_{[k]}(1) = \betaLassoTs$, $\hat \beta_{[k]}(0) = \betaLassoCs$,  $\beta_{[k]}(1) = \beta_{[k]\proj} (1)$, and $\beta_{[k]}(0) = \beta_{[k]\proj}(0)$, $k=1,\dots,K$. According to Corollary~\ref{cor::betaLasso} and the sparsity Assumption~\ref{assum::sparsity2}, we have
\begin{equation}
\label{eqn::consis-betaLassos}
|| \hat \beta_{[k]\Lasso}(a) -  \beta_{[k]\proj}(a)  ||_1 = O_P \left\{ s_{[k]} \bigg( \frac{M_{[k]n} \log p}{n} \bigg)^{1/2} \right\}= o_P(1), \quad a = 0,1.
\end{equation}
We have shown in the proof of Theorem~\ref{thm::Lasso} that
$$
 ||  \XkThat - \XkChat  ||_{\infty} = O_P\left\{   \bigg( \frac{\log p}{n} \bigg)^{1/2} \right\}.
$$
By H\"older inequality and \eqref{eqn::consis-betaLassos}, we have, for $a=0,1$ and $k=1,\dots,K$,
\begin{eqnarray}
&& | \surd{n} (  \XkThat -  \XkChat )^\T  \left\{  \hat \beta_{[k]\Lasso}(a) - \beta_{[k]\proj} (a) \right\}   | \nonumber \\
& \leq & \surd{n} || \XkThat -  \XkChat  ||_{\infty} \cdot  ||  \left\{  \hat \beta_{[k]\Lasso}(a)  - \beta_{[k]\proj}(a) \right\} ||_1 \nonumber \\
& = & O_P\left\{ \surd{n}  \bigg( \frac{\log p}{n} \bigg)^{1/2} \right\} \cdot O_P\left\{ s_{[k]} \bigg( \frac{M_{[k]n} \log p}{n} \bigg)^{1/2} \right\} \nonumber \\
& = & O_P\left( \frac{\surd{M_{[k]n}} s_{[k]} \log p}{ \surd{n} }  \right), \nonumber \\
& = & o_P(1), \nonumber
\end{eqnarray}
where the last equality is because of Assumption~\ref{assum::sparsity2}. The asymptotic normality of $\tauLassos$ and the consistency of the variance estimator follows from Theorem~\ref{thm::gen}.

Next, we compare the asymptotic variance of $\tauLassos$ and $\tauLasso$. Let $\Sigma_{[k] \bx Y(a)} = E [  \tilde \bx_i  \tilde Y_i(a)   \mid B_i = k ]$ be the stratum-specific covariance of $\bx_i$ and $Y_i(a)$. Similar to the proof of Theorem~\ref{thm::gen} with $r_{i,\gen}(a)$ replaced by $u_i(a)$, the difference between the asymptotic variance of $\tauLassos$ and $\taustr$ is
\begin{eqnarray}
\label{eqn::genvar-form-u}
\tilde \Delta & = & \{ \varsigma^2_{ u }(\pi) + \varsigma^2_{H u} \}  - \{  \varsigma^2_{Y}(\pi)  +   \varsigma^2_{HY}  \} \nonumber \\
 & = &   \frac{1}{\pi ( 1 - \pi ) } \sumk \pk (\beta_{[k] \proj}^{*} ) ^\T \Sigma_{[k] \bx \bx} (  \beta_{[k] \proj}^{*} ) \nonumber \\
 && - \frac{2}{ \pi ( 1 - \pi ) } \sumk \pk (\beta_{[k] \proj}^{*} ) ^\T \{ ( 1 - \pi )  \Sigma_{[k] \bx Y(1)} + \pi  \Sigma_{[k] \bx Y(0)} \} \nonumber \\
 & = &   \frac{1}{\pi ( 1 - \pi ) } \sumk \pk (\beta_{[k] \proj}^{*} ) ^\T \Sigma_{[k] \bx \bx} (  \beta_{[k] \proj}^{*} ) \nonumber \\
 & &  - \frac{2}{ \pi ( 1 - \pi ) } \sumk \pk (\beta_{[k] \proj}^{*} ) ^\T   \Sigma_{[k] \bx \bx}   \{ ( 1 - \pi )  \beta_{[k]\proj}(1) + \pi \beta_{[k]\proj}(0)  \} \nonumber \\
 & = &   \frac{1}{\pi ( 1 - \pi ) } \sumk \pk (\beta_{[k] \proj}^{*} ) ^\T \Sigma_{[k] \bx \bx} (  \beta_{[k] \proj}^{*} )   - \frac{2}{ \pi ( 1 - \pi ) } \sumk \pk (\beta_{[k] \proj}^{*} ) ^\T   \Sigma_{[k] \bx \bx}   (\beta_{[k] \proj}^{*} ) \nonumber \\
& = & -    \frac{1}{\pi ( 1 - \pi ) } \sumk \pk (\beta_{[k] \proj}^{*} ) ^\T \Sigma_{[k] \bx \bx} (  \beta_{[k] \proj}^{*} ), \nonumber
\end{eqnarray}
where the third equality is because  $\beta_{[k]\proj}(a) =  \Sigma_{[k] \bx \bx}^{-1} \Sigma_{[k] \bx Y(a)}$, and the fourth equality is because $  \beta_{[k] \proj}^{*} =  ( 1 - \pi )  \beta_{[k]\proj}(1) + \pi \beta_{[k]\proj}(0)  $.

According to Theorem~\ref{thm::Lasso}, the difference between the asymptotic variances of $\tauLasso$ and $\taustr$ is
$$
\Delta =  -  \frac{1}{\pi ( 1 - \pi ) }  ( \beta_{\proj}^{*} ) ^\T \Sigma_{ \tilde \bx \tilde \bx} ( \beta_{\proj}^{*} ).
$$
Therefore, the difference between the asymptotic variances of $\tauLassos$ and $\tauLasso$ is
\begin{eqnarray}
\Delta^* = \tilde \Delta - \Delta  =  -  \frac{1}{\pi ( 1 - \pi ) }  \left\{ \sumk \pk  (\beta_{[k] \proj}^{*} ) ^\T \Sigma_{[k] \bx \bx} (  \beta_{[k] \proj}^{*} )  - ( \beta_{\proj}^{*} ) ^\T \Sigma_{ \tilde \bx \tilde \bx} ( \beta_{\proj}^{*} ) \right\}, \nonumber
\end{eqnarray}
which is smaller than or equal to zero because $\tauLassos$ has the smallest asymptotic variance according to Theorem~\ref{thm::gen}.
\end{proof}

\section{Proof of lemmas}\label{sec::proof-lemma}

\subsection{Proof of Lemma~\ref{lem::concentration}}

\begin{proof}[Proof of Lemma~\ref{lem::concentration}]
We use the technique developed in \citet{Bugni2018} to prove Lemma~\ref{lem::concentration}. We will only prove  that
$$
|| \XkThat - E ( \bx_i \mid B_i = k ) ||_{\infty} = O_P\left\{   \bigg( \frac{\log p}{n} \bigg)^{1/2} \right\}, \quad ||  \XkThat - \XkT  ||_{\infty} = O_P\left\{   \bigg( \frac{\log p}{n} \bigg)^{1/2} \right\},
$$
as the proof for the counterpart of the control is similar. Let $A^{(n)} = \{ A_1,\dots, A_n \} $ and $B^{(n)}  = \{ B_1,\dots,B_n \} $. Note that by Assumptions~\ref{assum::Q}--\ref{assum::A}, $ \{ Y_i(1), Y_i(0), B_i, \bx_i \}_{i=1}^{n}$ are independent and identically distributed (i.i.d.), and $A^{(n)} $ are independent of $ \{ Y_i(1), Y_i(0), \bx_i \}_{i=1}^{n}$, conditional on $B^{(n)} $.  Then, conditional on $ \{   A^{(n)}, B^{(n)}  \} $, the distribution of $S_{\tilde \bx \tilde \bx  }(1) $ is the same as the distribution of the same quantity where units are ordered by strata and then ordered by $A_i=1$ first and $A_i=0$ second within each stratum. Thus, independently for each $k = 1,\dots,K$, and independent of $  \{   A^{(n)}, B^{(n)}  \}  $, let $\{  Y_i^k (1),  Y_i^k (0),  \bx_i^k \}$ be i.i.d. with marginal distribution being the same as the conditional distribution of $\{ Y_i(1), Y_i(0),  \bx_i \} $ given $ B_i = k$. Then, the conditional distribution of $  \XkThat - \XkT $ given $ \{   A^{(n)}, B^{(n)}  \} $ is the same as the distribution of
$$
\frac{1}{\nkt} \sum_{i=1}^{\nkt} \bx_i^k - \frac{1}{\nk} \sum_{i=1}^{\nk} \bx_i^k  =  \frac{1}{\nkt} \sum_{i=1}^{\nkt} \{  \bx_i^k - E (\bx_i^k  ) \} - \frac{1}{\nk} \sum_{i=1}^{\nk} \{ \bx_i^k  - E ( \bx_i^k  ) \} .
$$
Let $X_{ij}^k$ be the $j$th element of $\bx_i^k$. Under Assumption~\ref{assum::Q}, given $ B_i = k$,   $ \bx_i $ is uniformly bounded, and thus it is a sub-Gaussian random vector, then $\bx_i^k $ is also a sub-Gaussian random vector. Therefore, there exit constants $c_1$ and $c_2$ not depending on $n$, such that, for $t > 0$ and $j=1,\dots,p$,
\begin{equation}
P\left( \Big |    \frac{1}{m} \sum_{i=1}^{m} \left \{  X_{ij}^k - E( X_{ij}^k ) \right \} \Big | \geq t \right)  \leq c_1 \exp \{ - c_2 m t^2  \}. \nonumber
\end{equation}
Therefore,
\begin{eqnarray}
\label{eqn::concen1}
& = & P\left( \max_{j = 1, \dots, p} \Big |    \frac{1}{m} \sum_{i=1}^{m} \left\{  X_{ij}^k - E( X_{ij}^k ) \right\} \Big | \geq \sqrt{\frac{2 \log p }{c_2 m }}   \right)  \nonumber \\
& \leq & p  c_1 \exp \Big \{ - c_2 m  \frac{2 \log p }{c_2 m }  \Big \}  = c_1 \exp \{ - \log p \} \rightarrow 0.
\end{eqnarray}
Using the almost sure representation theorem, we can construct $\tilde n_{[k]1}$ (independent of $\{  \bx_i^k: i=1,\dots,n \}$) such that $\tilde n_{[k]1}/n$ has the same distribution as $\nkt /n$ and $\tilde n_{[k]1}/n \rightarrow \pi \pk$ almost surely (a.s.), thus,
\begin{eqnarray}
& & P\left( ||   \frac{1}{\nkt} \sum_{i=1}^{\nkt} \left\{  \bx_i^k - E ( \bx_i^k ) \right\} ||_{\infty} \geq \sqrt{ \frac{2}{c_2} } \sqrt{ \frac{ \log p }{ \nkt }} \ \ \Big |  \ \  A^{(n)}, B^{(n)}  \right)  \nonumber \\
& = & P\left( ||   \frac{1}{ n \frac{ \tilde n_{[k]1} }{n} } \sum_{i=1}^{n \frac{ \tilde n_{[k]1} }{n} } \left\{  \bx_i^k - E ( \bx_i^k ) \right\} ||_{\infty} \geq \sqrt{ \frac{2}{c_2} } \sqrt{ \frac{n}{\tilde n_{[k]1} } \frac{ \log p }{ n }}  \ \  \Big |  \ \  A^{(n)}, B^{(n)}  \right)  \nonumber \\
& = & E \left[  P\left( ||   \frac{1}{ n \frac{ \tilde n_{[k]1} }{n} } \sum_{i=1}^{n \frac{ \tilde n_{[k]1} }{n} }  \left\{  \bx_i^k - E ( \bx_i^k ) \right\}  ||_{\infty} \geq \sqrt{ \frac{2}{c_2} } \sqrt{ \frac{n}{\tilde n_{[k]1} } \frac{ \log p }{ n }} \ \  \Big |  \ \  A^{(n)}, B^{(n)} , \ \frac{ \tilde n_{[k]1} }{n} \right)   \right] \nonumber \\
& \rightarrow & 0, \nonumber
\end{eqnarray}
where the convergence follows from the dominated convergence theorem, $n (\tilde n_{[k]1}/n) \rightarrow \infty$ a.s., independence of $\tilde n_{[k]1}/n$ and $\{  \bx_i^k: i=1,\dots,n \}$, and \eqref{eqn::concen1}. As $\nkt / n  \xrightarrow{P} \pi \pk > 0$, we have
\begin{equation}\label{eqn::xik1}
|| \frac{1}{\nkt} \sum_{i=1}^{\nkt} \left\{  \bx_i^k - E ( \bx_i^k ) \right\} ||_{\infty} = O_P\left\{   \bigg( \frac{\log p}{n} \bigg)^{1/2} \right\}.
\end{equation}
Similar arguments yield
\begin{equation}\label{eqn::xik2}
|| \frac{1}{\nk} \sum_{i=1}^{\nk} \left\{ \bx_i^k  - E ( \bx_i^k ) \right\} ||_{\infty} = O_P\left(  \sqrt{\frac{\log p}{n}}  \right) .
\end{equation}
Making use of \eqref{eqn::xik1} and \eqref{eqn::xik2} yield
$$
||  \XkThat - \XkT ||_{\infty} = O_P\left\{   \bigg( \frac{\log p}{n} \bigg)^{1/2} \right\}.
$$
According to the above arguments,  the conditional distribution of $\XkThat - E ( \bx_i \mid B_i = k )$  given $ \{   A^{(n)}, B^{(n)}  \} $ is the same as the distribution of
$$
 \frac{1}{\nkt} \sum_{i=1}^{\nkt} \{  \bx_i^k - E (\bx_i^k  ) \}.
$$
Thus, by \eqref{eqn::xik1} and dominated convergence theorem, we have
$$
|| \XkThat - E ( \bx_i \mid B_i = k ) ||_{\infty} = O_P\left\{   \bigg( \frac{\log p}{n} \bigg)^{1/2} \right\}.
$$


\end{proof}

\subsection{Proof of Lemma~\ref{lem::stochastic}}
\begin{proof}
Recall that
$$
\r_i(a)  = Y_i(a) -    \bx_i ^\T \beta_{\proj}(a), \quad a = 0, 1.
$$
Then
\begin{eqnarray}
&& S_{\tilde \bx \r(1) } \nonumber \\
&=& \frac{1}{\nt} \sumk \sumik A_i ( \bx_i - \XkThat ) \{  \r_i(1) - \bar{\r}_{[k]1} \} \nonumber \\
&= & \frac{1}{\nt} \sumk   \sumik A_i  \left\{ \bx_i - E ( \bx_1 \mid B_1 = k)   \right\}  \{  \r_i(1) - \bar{\r}_{[k]1} \}  \nonumber \\
&&   - \frac{1}{\nt} \sumk \sumik A_i  \left\{  \XkThat  -  E ( \bx_1 \mid B_1 = k) \right\}  \{  \r_i(1) - \bar{\r}_{[k]1} \}  \nonumber \\
&= & \frac{1}{\nt} \sumk   \sumik A_i   \left\{ \bx_i - E ( \bx_1 \mid B_1 = k)   \right\}   \{  \r_i(1) - \bar{\r}_{[k]1} \},  \nonumber 
\end{eqnarray}
where the last equality is because
$$ \sumik A_i  \left\{ \XkThat - E ( \bx_1 \mid B_1 = k)  \right\}  \{  \r_i(1) - \bar{\r}_{[k]1} \} =  \left\{ \XkThat - E ( \bx_1 \mid B_1 = k)  \right\}  \sumik A_i    \{  \r_i(1) - \bar{\r}_{[k]1} \} = 0. $$
Let $\r^c_i(1) = \r_i(1) - E \{ \r_i(1) \mid B_i \}$ be the centered $\r_i(1)$. Then, $E\{ \r_i^c(1) \mid B_i = k \} = 0$, and
\begin{eqnarray}
\label{eqn::Sfirstterm}
&& S_{\tilde \bx \r(1) } \nonumber \\
& = & \frac{1}{\nt} \sumk   \sumik A_i   \left\{ \bx_i - E ( \bx_1 \mid B_1 = k)   \right\}   \{  \r^c_i(1) - \bar{\r}^c_{[k]1} \},  \nonumber \\
& = &  \frac{1}{\nt} \sumk   \sumik A_i   \left\{ \bx_i - E ( \bx_1 \mid B_1 = k)   \right\}    \r^c_i(1) -  \frac{1}{\nt} \sumk \nkt  \{ \XkThat - E (\bx_1 \mid B_1 = k) \}   \bar{\r}^c_{[k]1}. \nonumber \\
\end{eqnarray}

For the second term in  \eqref{eqn::Sfirstterm}, by Lemma~\ref{lem::concentration}, we have
$$
||  \XkThat - E ( \bx_1 \mid B_1 = k ) ||_{\infty}  = O_P\left( \sqrt{\frac{ \log p }{ n }} \right).
$$
Applying Lemma~\ref{lem::b3} to $\{ \r^c_i(1) \}$ within stratum $k$, together with $ E \{ \r^c_i(1) \mid B_i = k \} = 0 $,  we have
$$
\bar{\r}^c_{[k]1} = \frac{1}{\nkt} \sumik A_i  \r^c_i(1)  \xrightarrow{P} 0.
$$
Therefore, as the total number of strata $K$ is fixed, we have,
\begin{equation}
\label{eqn::stosecond}
|| \frac{1}{\nt} \sumk \nkt  \left\{ \XkThat - E ( \bx_1 \mid B_1 = k )  \right\}  \bar{\r}^c_{[k]1} ||_{\infty} = o_p \left( \sqrt{\frac{ \log p }{ n }} \right).
\end{equation}

For the first term in \eqref{eqn::Sfirstterm}, similar to the arguments in the proof of Lemma~\ref{lem::concentration}, conditional on $\{ A^{(n)}, B^{(n)} \}$,  it has the same distribution as
$$
\sumk \frac{\nkt}{\nt} \frac{1}{\nkt} \sum_{i=1}^{\nkt}  \left\{  \bx_i^k - E( \bx_i^k )  \right\} \r_i^k(1),
$$
where $\{ \bx_i^k, \r_i^k(1) \}$ are i.i.d. with marginal distribution being equal to the distribution of $\{ \bx_i, \r^c_i(1) \}  | B_i = k$,  independent for $k=1,\dots,K$, and independent of $\{ A^{(n)}, B^{(n)} \}$. For any integer $m \geq 1$,  using Nemirovski's inequality, there exists a constant $c$ not depending on $n$, such that
\begin{eqnarray}
&& E\left[ || \frac{1}{m}  \sum_{i=1}^{m}  \big\{  \bx_i^k - E( \bx_i^k ) \big\}  \r_i^k(1) - E \left[  \big\{  \bx_i^k - E( \bx_i^k )  \big\} \r_i^k(1) \right]  ||_{\infty}^2  \right] \nonumber \\
& \leq &\frac{ c \log p }{ m^2 }  \sum_{i=1}^{m} E \left[  ||  \big\{  \bx_i^k - E( \bx_i^k )  \big\}  \r_i^k(1)  - E \left[  \big\{  \bx_i^k - E( \bx_i^k )  \big\} \r_i^k(1) \right]   ||_{\infty}^2  \right] \nonumber \\
& \leq & \frac{ 4 c \log p }{ m^2 }  \sum_{i=1}^{m} E \left[  ||  \big\{  \bx_i^k - E( \bx_i^k )  \big\}  \r_i^k(1)  ||_{\infty}^2  \right] \nonumber \\
& \leq & \frac{16 M^2 c \log p }{ m^2 } \sum_{i=1}^{m} E   \big\{ \r_i^k(1) \big\}^2  \nonumber \\
& \leq & \frac{ 16 M^2 c \log p }{ m }  E \big\{  \r_1^k(1) \big\}^2 \nonumber \\
& \leq & c_{\lambda}  \frac{\log p}{m},
\end{eqnarray}
where
$$
c_{\lambda} = 16 M^2 c  \cdot \max_{k=1,\dots,K;\ a=0,1}  E \{ \r_1^k(a) \}^2 ,
$$ and the third inequality is because of Assumption~\ref{assum::Q} that $ || \bx_i ||_\infty \leq M$. Since $r_{i}(1) \in \mathcal{L}_2$, then, $ \r_1^k(1) \in \mathcal{L}_2$. Therefore, $c_{\lambda}$  is a constant not depending on $n$. Using Markov inequality, for any sequence $M_n$ tending to infinity,
\begin{eqnarray}
&& P\left(   || \frac{1}{m}  \sum_{i=1}^{m}  \bx_i^k   \r_i^k(1) - E \{  \bx_i^k \r_i^k(1)  \}  ||_{\infty} \geq  \sqrt{ c_{\lambda} M_n  \frac{\log p }{ m } }  \right) \nonumber \\
& \leq &  \frac{m}{c_{\lambda} M_n \log p} E \left[ || \frac{1}{m}  \sum_{i=1}^{m} \bx_i^k   \r_i^k(1) - E \{ \bx_i^k \r_i^k(1) \}  ||_{\infty}^2  \right] \nonumber \\
& \leq & \frac{1}{M_n}. \nonumber
\end{eqnarray}
Using similar arguments as those in the proof of Lemma~\ref{lem::concentration}, we have, for any $k=1,\dots,K$,
\begin{eqnarray}
\label{eqn::stofive}
&& P\left(     ||  \frac{1}{\nkt}  \sumik A_i  \left\{ \bx_i - E ( \bx_1 \mid B_1 = k)   \right\}   \r^c_i(1)   -  E \left[\left\{ \bx_i - E ( \bx_1 \mid B_1 = k)   \right\} \r^c_i(1) \mid B_i = k  \right]   ||_{\infty} \right. \nonumber \\
&& \quad \quad  \left.  \geq  \sqrt{   \frac{ c_{\lambda} M_n \log p }{ \nkt } }  \right) \leq \frac{1}{M_n}. \nonumber
\end{eqnarray}
By the property of projection,
$$
E\left[ \left\{  \bx_i  - E( \bx_i \mid B_i ) \right\} \r^c_i(1) \right] =  0 = \sumk \pk E \left[   \left\{ \bx_i - E ( \bx_i \mid B_i = k)   \right\}   \r^c_i(1)  \mid B_i = k \right] .
$$
Therefore,
\begin{eqnarray}
\label{eqn::stosix}
&& P\left( \Big | \Big |  \sumk \pk \frac{1}{\nkt}  \sumik A_i  \left\{ \bx_i - E ( \bx_1 \mid B_1 = k)   \right\}  \r^c_i(1) \Big | \Big |_{\infty}  \leq  \sumk \pk  \sqrt{ \frac{ c_{\lambda} M_n \log p }{ \nkt } } \right) \nonumber \\
& = &  P\left( \Big | \Big |  \sumk \pk \bigg( \frac{1}{\nkt}  \sumik A_i  \left\{ \bx_i - E ( \bx_1 \mid B_1 = k)   \right\}  \r^c_i(1)  -  E \left[   \left\{ \bx_i - E ( \bx_i \mid B_i = k)   \right\}   \r^c_i(1)  \mid B_i = k \right] \bigg) \Big | \Big |_{\infty} \right.   \nonumber \\
&& \quad \quad \left. \leq  \sumk \pk  \sqrt{ \frac{ c_{\lambda} M_n \log p }{ \nkt } } \right) \nonumber \\
& \geq & 1 - K \cdot \max_{k=1,\dots,K} P\left(     ||  \frac{1}{\nkt}  \sumik A_i  \left\{ \bx_i - E ( \bx_1 \mid B_1 = k)   \right\}   \r^c_i(1)   -  \right. \nonumber \\
&&\quad \quad \quad \quad \quad \quad \quad \quad \quad \quad  \left. E \left[\left\{ \bx_i - E ( \bx_1 \mid B_1 = k)   \right\} \r^c_i(1) \mid B_i = k  \right]   ||_{\infty} \geq  \sqrt{   \frac{ c_{\lambda} M_n \log p }{ \nkt } }  \right) \nonumber \\
&\geq & 1 - \frac{K}{M_n}.
\end{eqnarray}
Furthermore,
\begin{eqnarray}
& & || \sumk  \Big(  \frac{\nkt}{\nt}  - \pk \Big) \frac{1}{\nkt} \sumik A_i   \left\{ \bx_i - E ( \bx_1 \mid B_1 = k)   \right\}   \r^c_i(1)    ||_{\infty} \nonumber \\
& \leq &K \cdot \max_{k=1,\dots,K}  \Big|  \frac{\nkt}{\nt}  - \pk  \Big |  \cdot ||    \frac{1}{\nkt} \sumik A_i   \left\{ \bx_i - E ( \bx_1 \mid B_1 = k)   \right\}   \r^c_i(1)    ||_{\infty} \nonumber \\
& \leq &2 M K \cdot  \max_{k=1,\dots,K}  \Big|  \frac{\nkt}{\nt}  - \pk  \Big |  \cdot    \frac{1}{\nkt} \sumik A_i  |  \r^c_i(1) | , \nonumber
\end{eqnarray}
where the last inequality is because of the uniformly bounded assumption on  $\bx_i$. Since  $ \r_i(1) \in \mathcal{L}_2$, then  $ \r^c_i(1) \in \mathcal{L}_2$ and $ (1 / \nkt) \sumik A_i  |  \r^c_i(1) | = O_P(1)$ (by Lemma~\ref{lem::b3}).
Applying the asymptotic normality result of Proposition 1 in \citet{Ma2020Regression} to the  outcomes $D_i(1) = I_{B_i = k}$ and $D_i(0) = 0$, we have
$$
 \frac{\nkt}{\nt}  - \pk  =  O_p \left( {\frac{ 1 }{ \sqrt{n} }} \right) = o_p \left( \sqrt{\frac{\log p }{ n }} \right).
$$
Thus,
\begin{equation}
\label{eqn::stofour}
 || \sumk  \Big(  \frac{\nkt}{\nt}  - \pk \Big) \frac{1}{\nkt} \sumik A_i   \left\{ \bx_i - E ( \bx_1 \mid B_1 = k)   \right\}   \r^c_i(1)   ||_{\infty} = o_p \left( \sqrt{\frac{\log p }{ n }} \right).
\end{equation}
Combing~\eqref{eqn::Sfirstterm}, \eqref{eqn::stosecond}, \eqref{eqn::stosix} and \eqref{eqn::stofour}, when $n$ is large enough, it holds that
\begin{eqnarray}
P \left(  || S_{\tilde \bx \r(1) }  ||_{\infty} \geq 2  \sumk   \sqrt{ \frac{  c_{\lambda} \pk M_n  }{ \pi } }  \sqrt{ \frac{ \log p }{ n } }   \right ) \leq \frac{K}{M_n},
\end{eqnarray}
for any sequence $M_n \rightarrow \infty$. Thus, by Assumption~\ref{assum::lambda} on the tuning parameter $\lambda_1$, we have
$$
P( \mathcal{E}_1 ) = P( || S_{\tilde \bx \r(1) }  ||_{\infty} \leq \lambda_1 / 2 ) \geq 1 - \frac{K}{M_n} \rightarrow 1.
$$

\end{proof}

\subsection{Proof of Lemma~\ref{lem::RE}}

\begin{proof}
By the proof of Lemma~\ref{lem::concentration}, $ S_{\tilde \bx \tilde \bx  }(1) $ has the same distribution as
$$
 \sumk  \frac{\nkt}{\nt} \frac{1}{\nkt} \sum_{i=1}^{\nkt} \Big(  \bx_i^k  -  \frac{1}{\nkt} \sum_{i=1}^{\nkt}  \bx_i^k \Big)  \Big(  \bx_i^k  -  \frac{1}{\nkt} \sum_{i=1}^{\nkt}  \bx_i^k \Big)^\T.
$$
According to  Theorems 1.6 and 3.1 in \citet{ZhouS2009} (see also Theorem 1 and Corollary 1 in \citet{Raskutti2010} for correlated Gaussian distributed covariates), under Assumptions~\ref{assum::sub-Gaussian} and \ref{assum::sparsity}, there exits a constant $c_{\min}^k > 0$, such that when $m \rightarrow \infty$, for $\mathcal{C} = \{ \bh \in R^p:  || \bh_{S^c} ||_1 \leq 3 || \bh_{S} ||_1  \}$ and $\bh \in \mathcal{C}$,
\begin{eqnarray}
\label{eqn::RE1}
P\left(  \bh^\T  \Big[ \frac{1}{m} \sum_{i=1}^{m}  \Big(  \bx_i^k  -  \frac{1}{m} \sum_{i=1}^{m}  \bx_i^k \Big)  \Big(  \bx_i^k  -  \frac{1}{m} \sum_{i=1}^{m} \bx_i^k \Big)^\T  \Big]  \bh \geq   c_{\min}^k  || \bh ||_2^2 \right) \rightarrow 1.
\end{eqnarray}

Using the almost sure representation theorem, we can construct $\tilde n_{[k]1}$ (independent of $\{  Y_i^k (1), Y_i^k (0), \bx_i^k \}$) such that $\tilde n_{[k]1}/n$ has the same distribution as $\nkt /n$ and $\tilde n_{[k]1}/n \rightarrow \pi \pk$ a.s. Using the independence of $\{ A^{(n)}, B^{(n)} \}$ and $\{ Y_i^k (1),  Y_i^k (0), \bx_i^k \}$, we have, for $k=1,\dots,K$ and $\bh \in \mathcal{C}$,
\begin{eqnarray}
&& P\bigg(  \bh^\T  \Big[ \frac{1}{\nkt} \sum_{i=1}^{\nkt}  \Big( \bx_i^k  -  \frac{1}{\nkt} \sum_{i=1}^{\nkt} \bx_i^k \Big)  \Big( \bx_i^k  -  \frac{1}{\nkt} \sum_{i=1}^{\nkt} \bx_i^k \Big)^\T  \Big]  \bh \geq   c_{\min}^k  || \bh ||_2^2  \  \Big | \   A^{(n)}, B^{(n)}  \bigg) \nonumber \\
& = &  P\bigg(  \bh^\T  \Big[  \frac{1}{n \frac{ \tilde n_{[k]1}  }{n}} \sum_{i=1}^{n \frac{ \tilde n_{[k]1}  }{n}}  \Big( \bx_i^k  -  \frac{1}{n \frac{ \tilde n_{[k]1}  }{n} } \sum_{i=1}^{n \frac{ \tilde n_{[k]1}  }{n}} \bx_i^k \Big)  \Big( \bx_i^k  -  \frac{1}{n \frac{ \tilde n_{[k]1}  }{n} } \sum_{i=1}^{n \frac{ \tilde n_{[k]1}  }{n} } \bx_i^k \Big)^\T  \Big]  \bh \geq   c_{\min}^k  || \bh ||_2^2     \  \Big | \  A^{(n)}, B^{(n)} \bigg) \nonumber \\
& = & E \Bigg[  P \bigg(  \bh^\T  \Big[  \frac{1}{n \frac{ \tilde n_{[k]1}  }{n}} \sum_{i=1}^{n \frac{ \tilde n_{[k]1}  }{n}}  \Big( \bx_i^k  -  \frac{1}{n \frac{ \tilde n_{[k]1}  }{n} } \sum_{i=1}^{n \frac{ \tilde n_{[k]1}  }{n}} \bx_i^k \Big)  \Big( \bx_i^k  -  \frac{1}{n \frac{ \tilde n_{[k]1}  }{n} } \sum_{i=1}^{n \frac{ \tilde n_{[k]1}  }{n} } \bx_i^k \Big)^\T  \Big]  \bh \geq   c_{\min}^k  || \bh ||_2^2   \nonumber \\
& & \quad \quad \quad \  \Big |  \  A^{(n)}, B^{(n)}, \frac{\tilde n_{[k]1} }{n} \bigg)  \Bigg] \nonumber \\
& \rightarrow & 1, \nonumber
\end{eqnarray}
where the convergence follows from the dominated convergence theorem, $n (\tilde n_{[k]1}/n) \rightarrow \infty$ a.s., independence of $\tilde n_{[k]1}/n$ and $\{  Y_i^k (1), Y_i^k (0), \bx_i^k \}$, and \eqref{eqn::RE1}. Therefore, let $c_{\min} = \min\{c_1,\dots,c_K\} > 0$, we have
\begin{eqnarray}
&& P\Big(  \bh^\T S_{\tilde \bx \tilde \bx  }(1)   \bh \geq   c_{\min}  || \bh ||_2^2 \Big) \nonumber \\
& = & P\bigg(  \bh^\T \bigg[  \sumk  \frac{\nkt}{\nt} \frac{1}{\nkt} \sum_{i=1}^{\nkt} \Big( \bx_i^k  -  \frac{1}{\nkt} \sum_{i=1}^{\nkt} \bx_i^k \Big)  \Big( \bx_i^k  -  \frac{1}{\nkt} \sum_{i=1}^{\nkt} \bx_i^k \Big)^\T \bigg]  \bh \geq   c_{\min}  || \bh ||_2^2 \bigg) \nonumber \\
& \geq & P\bigg(  \bh^\T \bigg[  \frac{1}{\nkt} \sum_{i=1}^{\nkt} \Big( \bx_i^k  -  \frac{1}{\nkt} \sum_{i=1}^{\nkt} \bx_i^k \Big)  \Big( \bx_i^k  -  \frac{1}{\nkt} \sum_{i=1}^{\nkt} \bx_i^k \Big)^\T \bigg]  \bh \geq   c_{\min}^k  || \bh ||_2^2, \ k = 1,\dots,K \bigg)  \nonumber \\
& \rightarrow & 1.  \nonumber
\end{eqnarray}

\end{proof}


\section{Additional simulation results}

Tables~\ref{unequalsim} and \ref{unequalsim2} present the simulation results for sample size $n$ of $200$ and $500$ under unequal allocation ($\pi = 2/3$). The other simulation settings are the same as those described in the main text, and similar conclusions can be obtained as in the case of equal allocation.

\begin{table}[H]
	\centering
	\caption{Simulated bias, standard deviation, standard error, and coverage probability for different estimators and randomization methods under unequal allocation ($\pi=2/3$) and sample size $n = 200$.}\label{unequalsim}
	\vskip 5mm
	\begin{threeparttable}
		\setlength{\tabcolsep}{2pt}
\resizebox{\textwidth}{50mm}{
		\begin{tabular}{llcccccccccccccccccccccccc}
		\cline{1-20}
		&  & \multicolumn{6}{c}{Simple Randomization} & \multicolumn{6}{c}{Stratified Block Randomization} & \multicolumn{6}{c}{Minimization} \\ \cline{3-20}
		&
		&
		\multicolumn{1}{c}{Bias} &
		\multicolumn{1}{c}{SD} &
		\multicolumn{2}{c}{SE} &
        \multicolumn{2}{c}{CP} &
		\multicolumn{1}{c}{Bias} &
		\multicolumn{1}{c}{SD} &
		\multicolumn{2}{c}{SE} &
        \multicolumn{2}{c}{CP} &
		\multicolumn{1}{c}{Bias} &
		\multicolumn{1}{c}{SD} &
		\multicolumn{2}{c}{SE} &
        \multicolumn{2}{c}{CP}  \\ \cline{5-6} \cline{7-8} \cline{11-12} \cline{13-14} \cline{17-18} \cline{19-20}
		Model &
		Estimator &
		\multicolumn{1}{c}{} &
		\multicolumn{1}{c}{} &
		\multicolumn{1}{c}{unadj} &
		\multicolumn{1}{c}{adj} &
		\multicolumn{1}{c}{unadj} &
		\multicolumn{1}{c}{adj} &
		\multicolumn{1}{c}{} &
		\multicolumn{1}{c}{} &
		\multicolumn{1}{c}{unadj} &
		\multicolumn{1}{c}{adj} &
		\multicolumn{1}{c}{unadj} &
		\multicolumn{1}{c}{adj} &
		\multicolumn{1}{c}{} &
		\multicolumn{1}{c}{} &
		\multicolumn{1}{c}{unadj} &
		\multicolumn{1}{c}{adj} &
		\multicolumn{1}{c}{unadj} &
		\multicolumn{1}{c}{adj}   \\ \hline
  1&$\taustr$ & 0.10 & 5.83 & 5.78 & - & 0.95 & - & 0.06 & 5.87 & 5.76 & - & 0.95 & - & 0.00 & 5.71 & 5.73 & - & 0.95 & - \\
  &$\tauols$ & -0.01 & 1.84 & 1.77 & 1.79 & 0.93 & 0.94 & 0.03 & 1.80 & 1.76 & 1.78 & 0.94 & 0.94 & 0.01 & 1.80 & 1.75 & 1.77 & 0.94 & 0.94 \\
  &$\tauinter$ & 0.00 & 0.57 & 0.56 & 0.58 & 0.94 & 0.95 & 0.00 & 0.57 & 0.56 & 0.57 & 0.94 & 0.95 & -0.01 & 0.57 & 0.56 & 0.57 & 0.94 & 0.95 \\
  &$\tauLasso$ & 0.01 & 2.02 & 1.93 & 1.97 & 0.94 & 0.94 & 0.03 & 1.98 & 1.92 & 1.96 & 0.94 & 0.95 & 0.01 & 1.97 & 1.91 & 1.94 & 0.94 & 0.95 \\
  &$\tauLassos$ & 0.01 & 0.70 & 0.66 & 0.69 & 0.94 & 0.95 & 0.01 & 0.69 & 0.65 & 0.68 & 0.93 & 0.94 & -0.02 & 0.67 & 0.65 & 0.67 & 0.94 & 0.95 \\
  2&$\taustr$ & 0.04 & 3.08 & 3.05 & - & 0.94 & - & 0.01 & 3.06 & 3.05 & - & 0.95 & - & -0.01 & 3.05 & 3.06 & - & 0.95 & - \\
  &$\tauols$ & 0.08 & 2.54 & 2.50 & 2.54 & 0.94 & 0.95 & 0.04 & 2.53 & 2.50 & 2.54 & 0.95 & 0.95 & 0.06 & 2.48 & 2.51 & 2.54 & 0.95 & 0.95 \\
  &$\tauinter$ & 0.24 & 2.93 & 2.56 & 2.74 & 0.93 & 0.95 & 0.22 & 2.60 & 2.49 & 2.67 & 0.94 & 0.96 & 0.25 & 2.54 & 2.50 & 2.68 & 0.94 & 0.96 \\
  &$\tauLasso$ & 0.05 & 2.64 & 2.57 & 2.74 & 0.94 & 0.95 & 0.02 & 2.63 & 2.58 & 2.74 & 0.94 & 0.96 & 0.03 & 2.60 & 2.57 & 2.75 & 0.94 & 0.96 \\
  &$\tauLassos$ & 0.02 & 2.96 & 2.90 & 3.06 & 0.94 & 0.95 & -0.02 & 2.96 & 2.90 & 3.06 & 0.95 & 0.96 & -0.02 & 2.95 & 2.91 & 3.07 & 0.94 & 0.95 \\
  3&$\taustr$ & 0.05 & 2.00 & 1.86 & - & 0.93 & - & -0.04 & 1.97 & 1.87 & - & 0.93 & - & 0.03 & 1.98 & 1.86 & - & 0.93 & - \\
  &$\tauols$ & 0.00 & 0.30 & 0.29 & 0.30 & 0.94 & 0.95 & 0.00 & 0.29 & 0.28 & 0.29 & 0.94 & 0.95 & 0.00 & 0.30 & 0.28 & 0.29 & 0.94 & 0.95 \\
  &$\tauinter$ & -0.03 & 4.76 & 0.69 & 1.06 & 0.85 & 0.96 & 0.03 & 1.83 & 0.62 & 0.96 & 0.84 & 0.97 & 0.05 & 2.69 & 0.64 & 1.01 & 0.84 & 0.95 \\
  &$\tauLasso$ & 0.01 & 0.33 & 0.32 & 0.33 & 0.94 & 0.95 & 0.00 & 0.32 & 0.31 & 0.32 & 0.94 & 0.94 & 0.01 & 0.33 & 0.31 & 0.32 & 0.93 & 0.94 \\
  &$\tauLassos$ & 0.05 & 1.84 & 1.64 & 1.88 & 0.92 & 0.95 & -0.03 & 1.82 & 1.65 & 1.89 & 0.92 & 0.95 & 0.03 & 1.83 & 1.64 & 1.87 & 0.91 & 0.95 \\
   \hline
		\end{tabular}}
\begin{tablenotes}
\item Note: SD, standard deviation; SE, standard error; CP, coverage probability; \\
           \hspace{2cm} \hphantom{Note:} unadj, unadjusted variance estimator; adj, adjusted variance estimator; \\
           \hspace{2cm} \hphantom{Note:} -, not available.
\end{tablenotes}
\end{threeparttable}
\end{table}

\begin{table}[H]
	\centering
	\caption{Simulated bias, standard deviation, standard error, and coverage probability for different estimators and randomization methods under unequal allocation ($\pi=2/3$) and sample size $n = 500$.}\label{unequalsim2}
	\vskip 5mm
	\begin{threeparttable}
		\setlength{\tabcolsep}{2pt}
\resizebox{\textwidth}{50mm}{
		\begin{tabular}{llcccccccccccccccccccccccc}
		\cline{1-20}
		&  & \multicolumn{6}{c}{Simple Randomization} & \multicolumn{6}{c}{Stratified Block Randomization} & \multicolumn{6}{c}{Minimization} \\ \cline{3-20}
		&
		&
		\multicolumn{1}{c}{Bias} &
		\multicolumn{1}{c}{SD} &
		\multicolumn{2}{c}{SE} &
        \multicolumn{2}{c}{CP} &
		\multicolumn{1}{c}{Bias} &
		\multicolumn{1}{c}{SD} &
		\multicolumn{2}{c}{SE} &
        \multicolumn{2}{c}{CP} &
		\multicolumn{1}{c}{Bias} &
		\multicolumn{1}{c}{SD} &
		\multicolumn{2}{c}{SE} &
        \multicolumn{2}{c}{CP}  \\ \cline{5-6} \cline{7-8} \cline{11-12} \cline{13-14} \cline{17-18} \cline{19-20}
		Model &
		Estimator &
		\multicolumn{1}{c}{} &
		\multicolumn{1}{c}{} &
		\multicolumn{1}{c}{unadj} &
		\multicolumn{1}{c}{adj} &
		\multicolumn{1}{c}{unadj} &
		\multicolumn{1}{c}{adj} &
		\multicolumn{1}{c}{} &
		\multicolumn{1}{c}{} &
		\multicolumn{1}{c}{unadj} &
		\multicolumn{1}{c}{adj} &
		\multicolumn{1}{c}{unadj} &
		\multicolumn{1}{c}{adj} &
		\multicolumn{1}{c}{} &
		\multicolumn{1}{c}{} &
		\multicolumn{1}{c}{unadj} &
		\multicolumn{1}{c}{adj} &
		\multicolumn{1}{c}{unadj} &
		\multicolumn{1}{c}{adj}   \\ \hline
  1&$\taustr$ & -0.12 & 3.66 & 3.67 & - & 0.95 & - & -0.11 & 3.67 & 3.67 & - & 0.95 & - & 0.01 & 3.63 & 3.66 & - & 0.95 & - \\
  &$\tauols$ & -0.01 & 1.14 & 1.13 & 1.13 & 0.95 & 0.95 & 0.00 & 1.13 & 1.13 & 1.13 & 0.94 & 0.95 & -0.01 & 1.12 & 1.12 & 1.13 & 0.95 & 0.95 \\
  &$\tauinter$ & 0.00 & 0.36 & 0.36 & 0.36 & 0.95 & 0.95 & 0.00 & 0.36 & 0.36 & 0.36 & 0.95 & 0.95 & 0.00 & 0.36 & 0.36 & 0.36 & 0.95 & 0.95 \\
  &$\tauLasso$ & -0.02 & 1.19 & 1.18 & 1.19 & 0.95 & 0.95 & -0.01 & 1.19 & 1.18 & 1.18 & 0.95 & 0.95 & -0.01 & 1.17 & 1.17 & 1.18 & 0.95 & 0.95 \\
  &$\tauLassos$ & -0.01 & 0.39 & 0.39 & 0.39 & 0.95 & 0.95 & -0.01 & 0.39 & 0.39 & 0.39 & 0.95 & 0.95 & 0.00 & 0.39 & 0.39 & 0.39 & 0.95 & 0.95 \\
  2&$\taustr$ & -0.01 & 1.96 & 1.94 & - & 0.95 & - & 0.02 & 1.95 & 1.94 & - & 0.95 & - & -0.03 & 1.96 & 1.94 & - & 0.95 & - \\
  &$\tauols$ & -0.01 & 1.62 & 1.60 & 1.61 & 0.94 & 0.95 & 0.03 & 1.61 & 1.60 & 1.61 & 0.95 & 0.95 & 0.01 & 1.60 & 1.60 & 1.61 & 0.95 & 0.95 \\
  &$\tauinter$ & 0.06 & 1.61 & 1.58 & 1.62 & 0.95 & 0.95 & 0.09 & 1.60 & 1.58 & 1.62 & 0.95 & 0.95 & 0.07 & 1.59 & 1.58 & 1.62 & 0.95 & 0.95 \\
  &$\tauLasso$ & -0.02 & 1.64 & 1.62 & 1.65 & 0.94 & 0.95 & 0.02 & 1.63 & 1.62 & 1.65 & 0.95 & 0.96 & -0.01 & 1.63 & 1.62 & 1.65 & 0.95 & 0.95 \\
  &$\tauLassos$ & -0.01 & 1.73 & 1.68 & 1.80 & 0.94 & 0.96 & 0.03 & 1.73 & 1.68 & 1.79 & 0.95 & 0.96 & -0.01 & 1.72 & 1.68 & 1.80 & 0.94 & 0.96 \\
  3&$\taustr$ & 0.01 & 1.25 & 1.22 & - & 0.94 & - & 0.03 & 1.23 & 1.22 & - & 0.95 & - & 0.02 & 1.26 & 1.22 & - & 0.94 & - \\
  &$\tauols$ & 0.00 & 0.18 & 0.18 & 0.18 & 0.96 & 0.96 & 0.00 & 0.18 & 0.18 & 0.18 & 0.95 & 0.95 & 0.00 & 0.18 & 0.18 & 0.18 & 0.94 & 0.95 \\
  &$\tauinter$ & 0.00 & 0.81 & 0.25 & 0.32 & 0.91 & 0.96 & -0.01 & 0.64 & 0.22 & 0.28 & 0.91 & 0.96 & -0.01 & 0.73 & 0.23 & 0.30 & 0.91 & 0.95 \\
  &$\tauLasso$ & 0.00 & 0.19 & 0.20 & 0.20 & 0.95 & 0.95 & 0.00 & 0.20 & 0.19 & 0.19 & 0.95 & 0.95 & 0.00 & 0.20 & 0.19 & 0.19 & 0.94 & 0.95 \\
  &$\tauLassos$ & 0.00 & 0.74 & 0.65 & 0.74 & 0.91 & 0.95 & 0.02 & 0.70 & 0.63 & 0.71 & 0.92 & 0.95 & 0.01 & 0.73 & 0.63 & 0.71 & 0.91 & 0.94 \\
   \hline
		\end{tabular}}
\begin{tablenotes}
\item Note: SD, standard deviation; SE, standard error; CP, coverage probability; \\
           \hspace{2cm} \hphantom{Note:} unadj, unadjusted variance estimator; adj, adjusted variance estimator; \\
           \hspace{2cm} \hphantom{Note:} -, not available.
\end{tablenotes}
\end{threeparttable}
\end{table}

\section{Synthetic data of Nefazodone CBASP trial}
To generate the synthetic data, we first fit non-parametric splines using the function \textbf{bigssa} in the R package \textbf{bigspline} with treatment indicator (1 for combination and 0 for Nefazodone), stratification covariate GENDER and eight baseline covariates: AGE, HAMA\_SOMATI, HAMD17, HAMD24, HAMD\_COGIN, Mstatus2, NDE and TreatPD. The baseline covariates are detailed in Table~\ref{tab::description}. The fitted model can be loaded from the files spline0.RData and spline1.RData.

\begin{table}[H]
\centering
\caption{Description of baseline covariates}\label{tab::description}
\begin{tabular}{l|l}
\hline
 Variable & Description \\
 \hline
 AGE  & Age of patients in years \\
 GENDER & 1 female and 0 male \\
 HAMA\_SOMATI & HAMA somatic anxiety score \\
 HAMD17 & Total HAMD-17 score \\
 HAMD24  & Total HAMD-24 score \\
 HAMD\_COGIN & HAMD cognitive disturbance score  \\
 Mstatus2 & Marriage status: 1 if married or living with someone and 0 otherwise \\
 NDE & Number of depressive episode \\
 TreatPD & Treated past depression: 1 yes and 0 no \\
 \hline
\end{tabular}
\begin{tablenotes}
\item Note: HAMD, Hamilton Depression Scale; HAMA, Hamilton Anxiety Scale.
\end{tablenotes}
\end{table}

Then, we  implement simple randomization and stratified block randomization to obtain the treatment assignments for both  equal ($\pi=1/2$) and unequal ($\pi=2/3$) allocations. We use HAMD17 and AGE for OLS-adjusted treatment effect estimators, and take linear, quadratic, cubic, and interaction terms of continuous covariates, linear and interaction terms of binary covariates, and interaction terms of the above two sets of coordinates as the covariates ($p=135$) for Lasso-adjusted treatment effect estimators.

\end{document}